\documentclass{bmcart}

\usepackage{amsthm,amsmath}
\usepackage{amsmath,amssymb}
\RequirePackage{hyperref}
\usepackage[applemac]{inputenc} 
\usepackage{graphicx}
\usepackage{bm}
\usepackage{dcolumn}
\usepackage[mathscr]{eucal}
\usepackage{color}

\newcommand{\bea}{\begin{eqnarray}}	
\newcommand{\eea}{\end{eqnarray}}
\newcommand{\be}{\begin{equation}}	
\newcommand{\ee}{\end{equation}}

\startlocaldefs
\endlocaldefs

\begin{document}

\begin{frontmatter}

\begin{fmbox}
\dochead{Research}


\title{GEMs and amplitude bounds in the colored Boulatov model} 


\author[
   addressref={aff1},                   
   corref={aff1},                       
   email={f.caravelli@ucl.ac.uk}   
]{\fnm{Francesco} \snm{Caravelli}}


\address[id=aff1]{
  \orgname{QASER Lab, University College London}, 
  \street{Gower Street},                     %
  \city{London},                              
  \cny{UK}                                    
}


\end{fmbox}


\begin{abstractbox}
\begin{abstract} 
In this paper we construct a  methodology for separating the divergencies due to different topological manifolds dual to Feynman graphs  in colored group field theory. 
After having introduced the amplitude bounds using propagator cuts, we show how Graph-Encoded-Manifolds (GEM) techniques can be used in order to factorize divergencies related to different parts of the dual topologies of the Feynman graphs in the general case. We show the potential of the formalism in the case of 3-dimensional solid torii in the colored Boulatov model.

\end{abstract}




\end{abstractbox}
%

\end{frontmatter}

\section{Introduction}
Recently there has been a growth of interest in group field theories 
\cite{daniele,razjim,quantugeom2} and there are many reasons for this to happen, first of all because these are connected with spin foams\cite{sf}. Spin foams are a formalization of covariant time evolution in Loop Quantum Gravity, and group field theories can be seen as a formalization of their partition function expansion. In this sense, Group field theories (GFT) can be formalized as Quantum Field Theory over a group manifold and are a generalization of matrix models to higher dimensions
\cite{Brezin:1977sv}\cite{mm}. It is known that matrix models have a topological 
expansion in which the genus, the only topological 
invariant needed to characterize 2-surfaces, plays the role of the parameter 
of this expansion. 
Roughly speaking, a $n$-dimensional GFT has a vertex associated to 
an $n$-simplex and a propagator which glues the $n$-simplices along their $(n-1)$-dimensional boundary faces. 
The Feynman diagrams of a $n$-dimensional GFT, in their dual, can be interpreted as gluings of simplices and then have the interpretation of 
piecewise linear (PL) manifolds. 
However, generic GFT in 3 dimensions have the problem that the gluings are too 
arbitrary, in the sense that the generated simplicial 
complexes are not even pseudo-manifolds, since they present wrapping singularities\cite{sing}. As in ordinary $\phi^4$ theory, a 
3-dimensional GFT can generate a ``8'' diagram of which the dual has no obvious topological interpretation in the continuum limit. 
This phenomenon does not happen in 2-dimensions, and the hope is that these singularities can be removed in the continuum limit. For this reason a \textit{colored} 
version of group field theory (cGFT) was 
introduced in an arbitrary number of dimensions (see \cite{color,razjim} and references therein), not to mention a generalization of group field theory to tensor\cite{tensor}, for which a universality theorem has been proven in the case of independent fluctuations\cite{universality}.
The challenge in these models is to obtain a topological expansion as in the 
2-dimensional case \cite{sefu3,FreiGurOriti,sefu1,smerlakbom,sylv1}. It has been shown
in \cite{1on} that diagrams whose dual topology is the one of the sphere dominates the partition function in the limit of the cutoff being removed in an arbitrary number of dimensions. In order to achieve this result, techniques from the theory of crystallizations\cite{Pezzana,Lins} have been used.
Colored $n$-graphs are well known in mathematics as \textit{gems}: 
graph-encoded manifolds \cite{Pezzana,Lins}. 
In a previous paper \cite{mio} we used GEMs techniques in order to prove the orientability
of pseudo-manifolds generated in cGFT in any dimensions, based on the fact
that these generate only bipartite graphs. In this paper, we base our methodology on 
another theorem proved in the field of crystallizations, which puts in relation dipole-contractions and 
the connected sum of manifolds. 

In order to clarify further the background, we briefly give a review of the context. There is a class of graphs which plays a special role in cGFT. These are called melonic graphs. The simpler of these is a 2-node graph in which all the propagators are connected but two of them, which are of the same colour. Now, we can construct any type of melonic graph from this, by inserting a melonic graph inside each of the propagators in all possible combinations. It is easy to see that this type of graphs can be classified in terms of trees. It has been shown in \cite{1on} that this class of graphs, remarkably, dominate the partition function in the limit in which the cutoff goes to infinity, and in particular, their duals are always associated to spheres. In fact, fixed the number of nodes, a melonic graph has maximum divergence. Initiated by \cite{1on}, a series of studies \cite{gurau3,gurau4,sylv3,sylv4,jimmy} has elucidated the structure of the divergencies in many cases, and a class of renormalizable theories have been introduced in \cite{geloun,geloun2}. However, as noted in \cite{rivasseau}, and further studied in \cite{gurau2}, spheres are combinatorially not favoured to other topologies in the perturbative expansion. For this reason we believe it is important to have an understanding of the corrections to the path integral given by Feynman graphs whose duals correspond to topologies other than spheres, and techniques like those we will introduce in the body of this paper could be potentially useful.

We will try to address this in a framework of ``topological'' cuts. How do we separate the divergencies coming from spheres, and those coming from other topologies?
In other terms, how do we factorize these divergencies? A step towards this type of problem will be put forward in this paper, where
we will argue that performing the cuts strategically factorizes graphs whose duals are spheres. When these cuts are performed, the graphs combine into connected sums and thus
the remaining graphs is actually a well defined surgery operation in topology. 

For the case of 3-manifold, a theorem \cite{knemiln} ensures that that every compact and orientable 3-manifold can be decomposed uniquely into
the connected sum of prime manifolds. Although Group Field Theory generates only pseudo-manifolds, their orientability has been shown in \cite{mio}.
Thus, if we were able to eliminate pseuso-manifolds from the partition function, the techniques developed in this paper could be very useful.
In the following we will focus on 3-dimensional cGFT, in particular the colored Boulatov model, as the theorem we will make use of is based on 3-dimensional topology.

The paper is organized as follows: in section II we recall the colored Boulatov model and its
standard interpretation. In section III we develop the cuts strategy, main result of this paper, and in section IV we apply it to the connected sum of solid toriii as an example.
Conclusions follow.

\section{The colored Boulatov Model and 3-gems}
We now introduce the colored Boulatov model\cite{GFT, color}.
Consider a compact Lie group $G$, denote $h$ its elements, $e$ the unit element, and $\int dh$ the integral with respect to the Haar measure of the group.

In 3-dimensions, we introduce two fields, $\bar \psi^i$ and $\psi^i$,  $i=0,1,2,3$ be four couples of complex 
scalar (or Grassmann) fields over four copies of $G$, $\psi^i:G\times G\times G \times G 
\rightarrow \mathbb{C}$. In a generic number of dimensions $i=0,\cdots,n+1$ where $n$ is the number of dimensions, and
the $\psi$ and $\bar \psi$ are functions of $n$ copies of the group. We define $e$ as the identity element of the group and we denote $\delta^\Lambda(h)$, the 
regularized delta function over $G$ with some cutoff $\Lambda$ such that 
$\delta^\Lambda(e)$ is finite, but diverges when $\Lambda$ goes to infinity. A feasible regularization is given, 
for instance for the group $G=SU(2)$, by
\begin{equation}
  \delta^\Lambda(h) = \sum_{j=0}^{\Lambda} (2j+1) \chi^{j}(h)
\label{cutoff}
\end{equation}
where $\chi^j(h)$ is the character of $h$ in the representation $j$, and which preserves the composition properties. The path integral for the colored Boulatov model over $G$ is:
\bea\label{eq:part}
&Z(\lambda,\bar\lambda)= e^{-F(\lambda,\bar\lambda)} = \int \prod_{i=0}^3 d\mu_P(\bar \psi^i,\psi^i) \; e^{-S^{int}(\bar \psi^i,\psi^i)} \; ,
\eea
where the Gaussian measure $d \mu_P$, with $P$ being its covariance, is chosen such that:
$$\int \prod_{i=0}^4 d\mu_P(\bar \psi^i,\psi^i)=1\ ,$$
and:
\bea
P_{h_{0}h_{1}h_{2} ; h_{0}'h_{1}'h_{2}'} && = \int d\mu_P(\bar \psi^i,\psi^i) \; \bar\psi^i_{h_{0}h_{1}h_{2}} \psi^i_{h_{0}'h_{1}'h_{2}'} = \nonumber \\
&& = \int dh \;  \delta^{\Lambda}\bigl( h_{0} h (h_{0}')^{-1} \bigr) \delta^{\Lambda}\bigl( h_{1} h (h_{1}')^{-1} \bigr) \delta^{\Lambda}\bigl( h_{2} h (h_{2}')^{-1} \bigr) \; \nonumber  ,
\eea
The colored model has two interactions, a ``clockwise`` and an ``anti-clockwise``, and one is obtained from the other one by complex
conjugation in the internal group color, one for each face of the 3-simplex. 
We fix the notation for shortage of space, $\psi(h,p,q)=\psi_{hpq}$. There are two interaction terms:  

\begin{eqnarray}
 S^{int} &=  \frac{\lambda}{\sqrt{\delta^\Lambda (e) } }\int \prod_{i,j} dh_{i,j}\ \psi^0_{h_{03}h_{02}h_{01} }  \psi^1_{h_{10}h_{13}h_{12}} \psi^2_{h_{21}h_{20}h_{23}}  \psi^3_{h_{32}h_{31}h_{30}} + \nonumber \\
& \frac{\bar\lambda}{\sqrt{\delta^\Lambda(e)}} \int  \prod_{i,j} dh_{i,j}\ \bar \psi^0_{h^{03}h^{02}h^{01} }  \bar \psi^1_{h^{10}h^{13}h^{12} }  \bar \psi^2_{ h^{21}h^{20}h^{23} }  \bar \psi^3_{h^{32}h^{31}h^{30} } 
\label{eq:interaction}
\end{eqnarray}
where $h_{ij}=h_{ji}$. In order to make the notation clearer, we call ``red'' the vertex involving the $\psi$'s and
``black'' the one involving the $\bar \psi$'s. Thus any line coming out of a cGFT vertex has a color $i$. 

The group elements $h_{ij}$ in eq. (\ref{eq:interaction}) are associated to a field, and glue two vertices with opposite orientation. 

In the body of this paper we will consider only vacuum graphs, i.e. all the vertices all the graphs are 4-valent (no open lines) and we will only deal with connected graphs.
The strands of a vacuum cGFT graph $\Gamma$ have a natural orientation given by the fact that only vertices of opposite orientations can be glued. It is easy to see that a vacuum cGFT graph must have the same number of black and red vertices. 

For any graph $\Gamma$, consider the set $\mathscr V_\Gamma$ of its vertices, $|\mathscr V_\Gamma|=2n$, the set $\mathscr L_\Gamma$ as its edges, and we define as \textit{faces} $\mathscr F_\Gamma$ (not to be confused with the faces of the tetrahedron!), as any closed line in the Feynman graph of a GFT. 

A generic vacuum Feynman amplitude of the theory can be written as:
\bea\label{eq:ampli}
 \mathscr A  = \frac{(\lambda\bar\lambda)^{\frac{n}{2}}} { [\delta^\Lambda (e)]^{n}} 
\int \prod_{l\in \mathscr L_\Gamma} dh_{l} 
\prod_{f\in \mathscr F_{\Gamma}} \delta^\Lambda_{f}(\prod_{l_0 \in f }^{\rightarrow} h_{l_0}^{\sigma^{l_0\ |\ f}} )\; ,
\eea
where the notation $l_0\in f$ means that the line $\ell$ belongs to the face $f$ and $\sigma^{l_0\ |\ f}=1$ (resp. $-1$)
if the orientations of $\ell$ and $f$ coincide (resp. are opposite). In the following we will assume that an orientation is fixed. The $\delta^\Lambda$ functions are invariant
under cyclic permutations and conjugation of their arguments. Hence the amplitude of a graph does not depend on the
orientation of the faces. 

We now recall few facts about 3 graph-embedded pseudo-manifold, which are strictly related to the colored Boulatov model and enlight the topological properties of the Feynman graph duals. 
Let $\Gamma$ be a finite, edge-colored graph.  A $k$-\textit{bubble} in the spinfoam literature) of $\Gamma$, $k\in \textbf{N}$ is a connected component of subgraph of $\Gamma$ induced by $k$ colors. 

These graphs represent a piecewice linear pseudo-complex in the following sense \cite{surveyger}. 
Let $\Gamma$ be a 4-regular graph, $\gamma$ its coloring. To a couple $(\Gamma,\gamma)_{n+1}$ there is an associated 
\textit{pseudo}-complex $K(\Gamma)$ given by the following construction. Take an $n$-simplex $\sigma^n$ for each $\mathscr V_\Gamma$ and label its vertices $\Delta_n$. 
If $x_i$,$y_j$ in $\mathscr V_\Gamma$ are joined by an edge, then we glue also the related $(n-1)$-simplices. In the case of a tetrahedron, this means gluing the faces of tetrahedron, which are triangles, of the same color. We denote with $|\Gamma|$ the pseudo-complex associated with the colored graph $\Gamma$. It is the color makes this procedure unambiguous, as there is only one color for each face of the tetrahedron. This is the same interpretation given to connecting vertices (of opposite orientation) in a (colored) n-dimensional GFT.  This is also the reason why colored group field theories have a clear interpretation as orientable pseudo manifolds \cite{mio}.

Thus, given the above description of a pseudo-complex, cGFT are associated to complexes in the following sense: a vertex can be seen as the dual of a tetrahedron and 
its propagators represent gluings of the triangles which form the tetrahedron of the same color. 

\begin{figure}[htb]
\begin{center}
\includegraphics[scale=0.4]{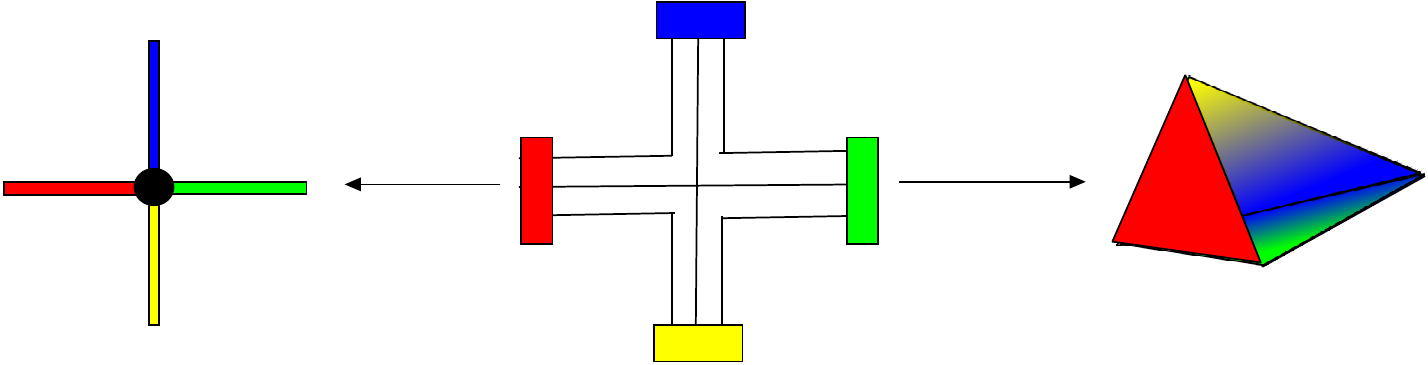}\hspace{0.5cm} 
\includegraphics[scale=0.4]{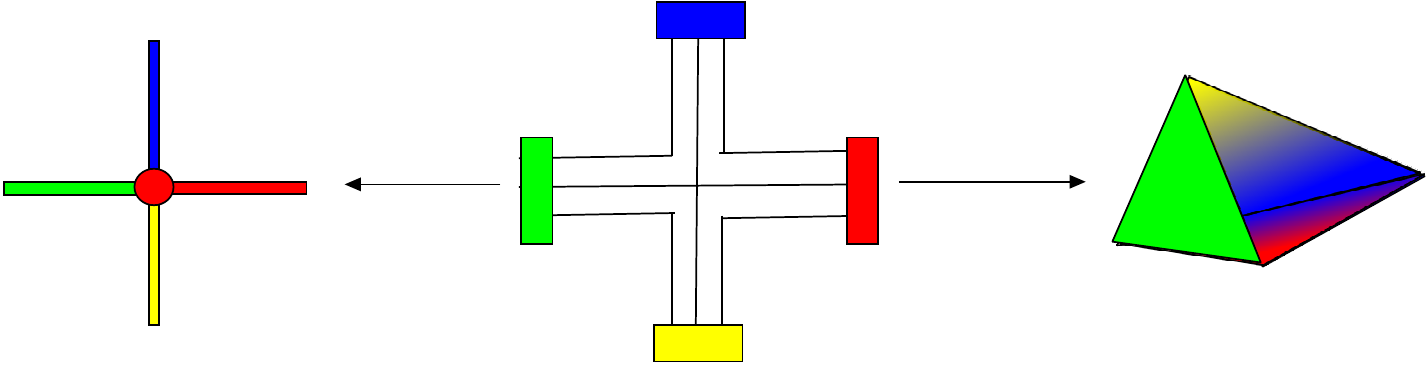}\hspace{0.5cm} 
\caption{Colored GFT red and black vertices.}
\label{fig:vertex}
\end{center}
\end{figure}
\begin{figure}[htb]
\begin{center}
\includegraphics[scale=0.4]{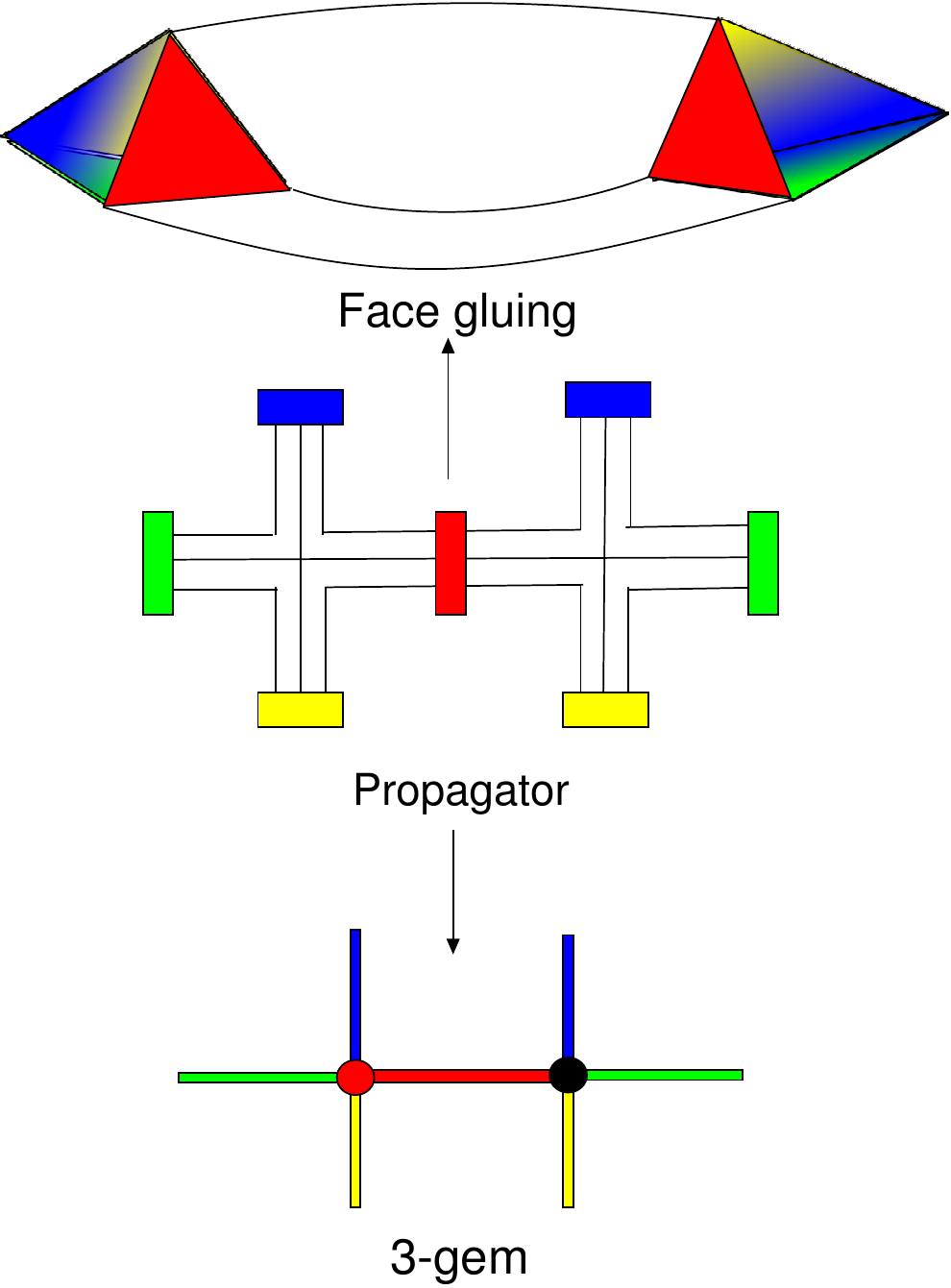}\hspace{0.5cm}
\caption{Gluing through propagator.}
\label{fig:prop}
\end{center}
\end{figure}

Each propagator is actually decomposed into three {\it parallel}  strands
which are associated to the three arguments (group elements) of the fields, i.e. the 1-dimensional elements of the 1-skeleton of the tetrahedron which bound every face. A line represents the gluing of 
two tetrahedra (of opposite orientations) along triangles of the same color as in Fig. (\ref{fig:prop}).

Another operation which we need to introduce is the connected sum of two manifold, denoted by the symbol \#. The connected sum of two manifolds $M_1$ and $M_2$, is given by 
$M \backsimeq M_1\ \#\ M_2$.  If both manifolds are oriented, there is a unique connected sum operation. Consider two balls, one in $M_1$ and one $M_2$. Carve the two balls, and glue their boundaries with opposite orientation. The points where the two balls have been carved denotes the two points which characterize the operation, which is then $\#_{p_1, p_2}$. The result is unique up to homeomorphisms, and thus when the base points will be omitted the result won't be ambiguous. 

\begin{figure}[htb]
\center
\includegraphics[scale=0.6]{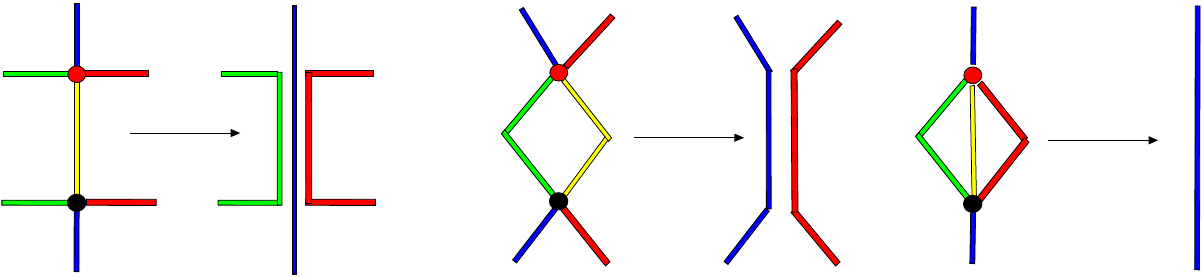}
\caption{Dipole contraction and creation on a 4-regular 4-edge colored graph of 1-, 2- and 3- dipoles respectively.}
\label{fusionm} 
\end{figure}

We now introduce dipole contractions and creations. Dipole contraction is an operation performed between two nodes of a Feynman colored graph, as described in Fig. \ref{fusionm} for the case of 4-colored graph.

We call two dipoles degenerate if they belong to the same bubble, and non-degenerate otherwise.

We now introduce the main theorem which will be central to our analysis hereafter \cite{Lins}.\\\ \\
\textbf{Theorem} (dipole contraction) Let $u$ and $v$ be vertices in a 4-graph connected by an m-dipole, and $\Gamma_1$ and $\Gamma_2$ two colored 4-graphs. If $S^n$ is the n-dimensional sphere, we have:
\begin{list}{}{}
\item a) $|\Gamma_1\ \#_{uv}\ \Gamma_2|\backsimeq |\Gamma_1|\ \#\ |\Gamma_2|$:
\item b) If $u$ and $v$ are not contained in the same 3-bubble of $\Gamma_1$ and $\Gamma_2$ is obtained from $\Gamma_1$ by fusion on $u$ and $v$, then $|\Gamma_2| \backsimeq |\Gamma_1|\ \#\ S^1 \otimes S^2$;
\item c) If $\Gamma_1$ is obtained from $\Gamma_2$ by removing a \textit{degenerate} 2-dipole, then $|\Gamma_2| \backsimeq |\Gamma_1|\ \#\ S^1 \otimes S^2$;
\item d) If $\Gamma_1$ is obtained from $\Gamma_2$ by removing a degenerate 1-dipole then one of the following cases hold:
\end{list}
\begin{list}{}{}

\item \ \ i) $|\Gamma_1|\backsimeq|\Gamma_2|$ if $u$ and $v$ are contained in a 2-bubble;
\item \ \ ii) $|\Gamma_1|\backsimeq|\Gamma_2|\ \#\ S^1 \otimes S^2$ if $u$ and $v$ are contained in 2  different 2-bubbles;
\item \ \ iii) $|\Gamma_1|\backsimeq|\Gamma_2|\ \#\ S^1 \otimes S^2\ \#\ S^1 \otimes S^2$ if three such 2-bubbles and $\Gamma_2$ are connected;
\item \ \ iv)  $|\Gamma_1|\backsimeq  |\Gamma_1'|\ \#\ |\Gamma_2'|\ \#\ S^1 \otimes S^2$ if three such 2-bubbles and $\Gamma_2$ has two connected components $\Gamma_1'$ and $\Gamma_2'$.
\end{list}
depending on whether the endpoints $u$ and $v$ of the dipole are both contained on exactly (i) one 2-bubble with colors other than that of edge \{$u$,$v$\}; (ii) two such 2-bubbles; (iii) three such 2-bubbles and $\Gamma'$ is connected; 
or (iv) three such 2-bubbles and $\Gamma'$ has two connected components $\Gamma_1'$ and $\Gamma_2'$.\\\ \\

This theorem is fundamental in order to have a clear understanding of topology associated with graphs generated by a cGFT and it will be extensively used in the rest of the paper, as it also makes clear the role of the connected sum operation in relation to dipole moves.

\begin{figure}[htb]
\center

 \includegraphics[scale=0.5]{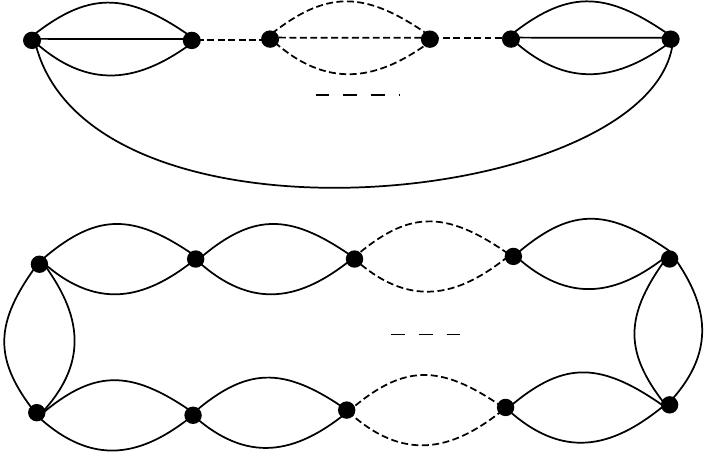}
 
 \caption{Connected sum of spheres $S^3$. The graph on top has a divergence degree proportional to the number of vertices, while the graph on the bottom has a divergence degree independent from the number of vertices.}
\label{connsumsph}
\end{figure}

Let us first make some remarks. The first thing we note is that the connected sum of spheres is given by only 2- and 3-dipoles. Thus, by a subsequent contraction of these, we can always return to a graph with 2 nodes, and according to the theorem above, this will be homeomorphic to the sphere.
This is true, for instance, for the graphs in Fig. \ref{connsumsph}. As we will see later, however, the divergencies of the two graph shown in Fig. \ref{connsumsph} are different.
these are two expamplicative cases  The reason is that
2-dipoles in ``parallel'' generate divergencies, while 2-dipoles in series do not. This is the reason why 2-dipoles are quite bad to handle, while 3-dipoles are not (they always produce a divergence and contract to a line).
Moreover this is the main difference between 2-dimensional matrix models and their generalizations in higher dimensions. In the case of 2-dimensional matrix models adding a sphere does not produce any divergence, and the reason why 3-dipoles. The reason is due to the fact that
$S_2\ \#\ S_2$ has still the same genus and thus in the 't Hooft limit the scale in the same fashion. The 3-dimensional case is, as we just have remarked, quite different. We would like to make the point that genus zero 
manifolds in 3-dimensions are \textit{unique}: adding or not a sphere does not change the topology at all. Moreover, amplitudes are indipendent under the change of orientation of the manifold, as this is the same as acting through the color symmetry of the group field theory \cite{mio}. 

\begin{figure}[htb]
\center
\includegraphics[scale=0.5]{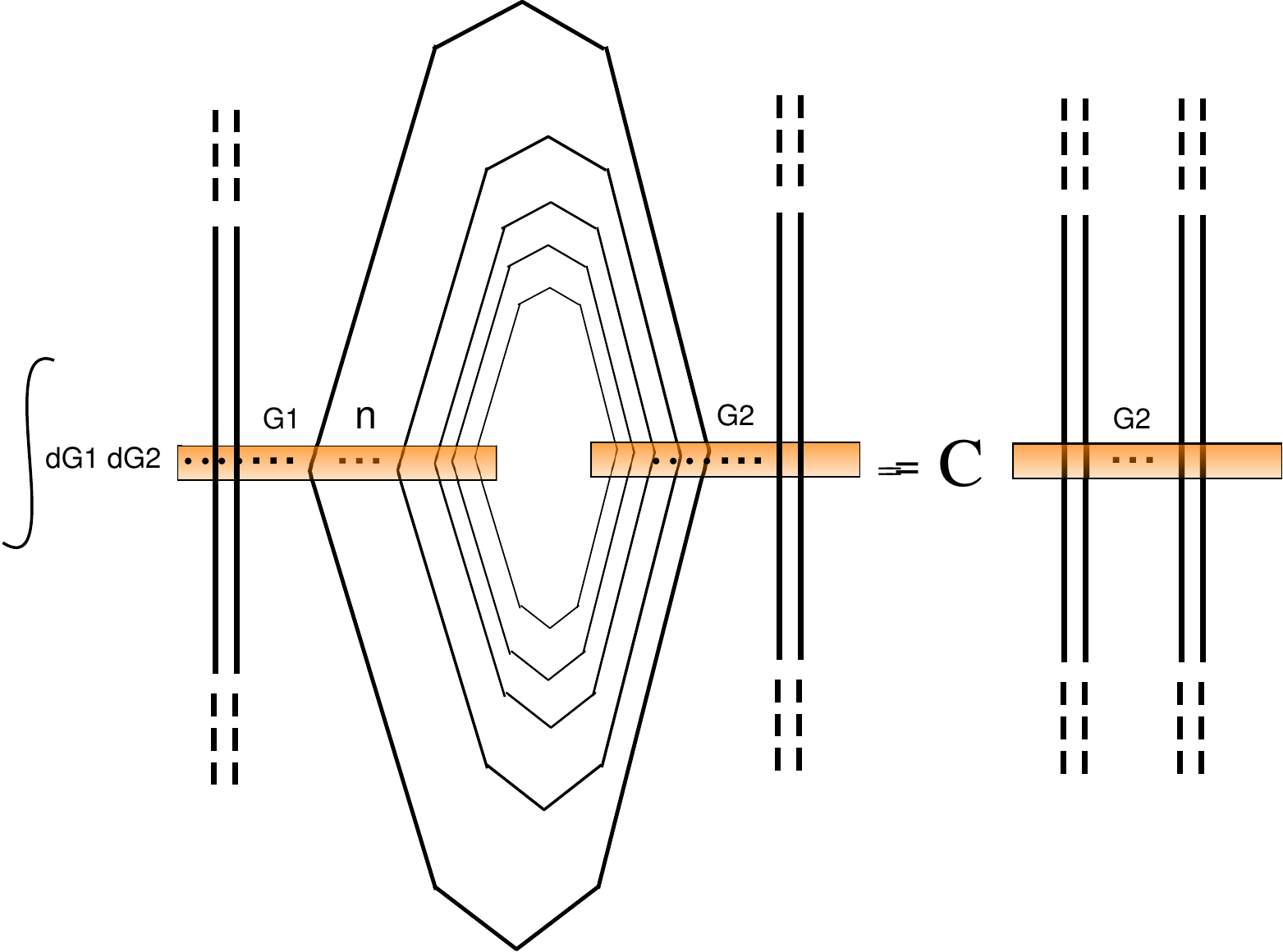}
\caption{The integration rule for the reduction of concentric strands. The constant $C$ is infinite and, regularized, is $[\delta^\Lambda(e)]^{n-1}$.}
\label{div}
\end{figure}

\begin{figure}[htb]
\center
\includegraphics[scale=0.4]{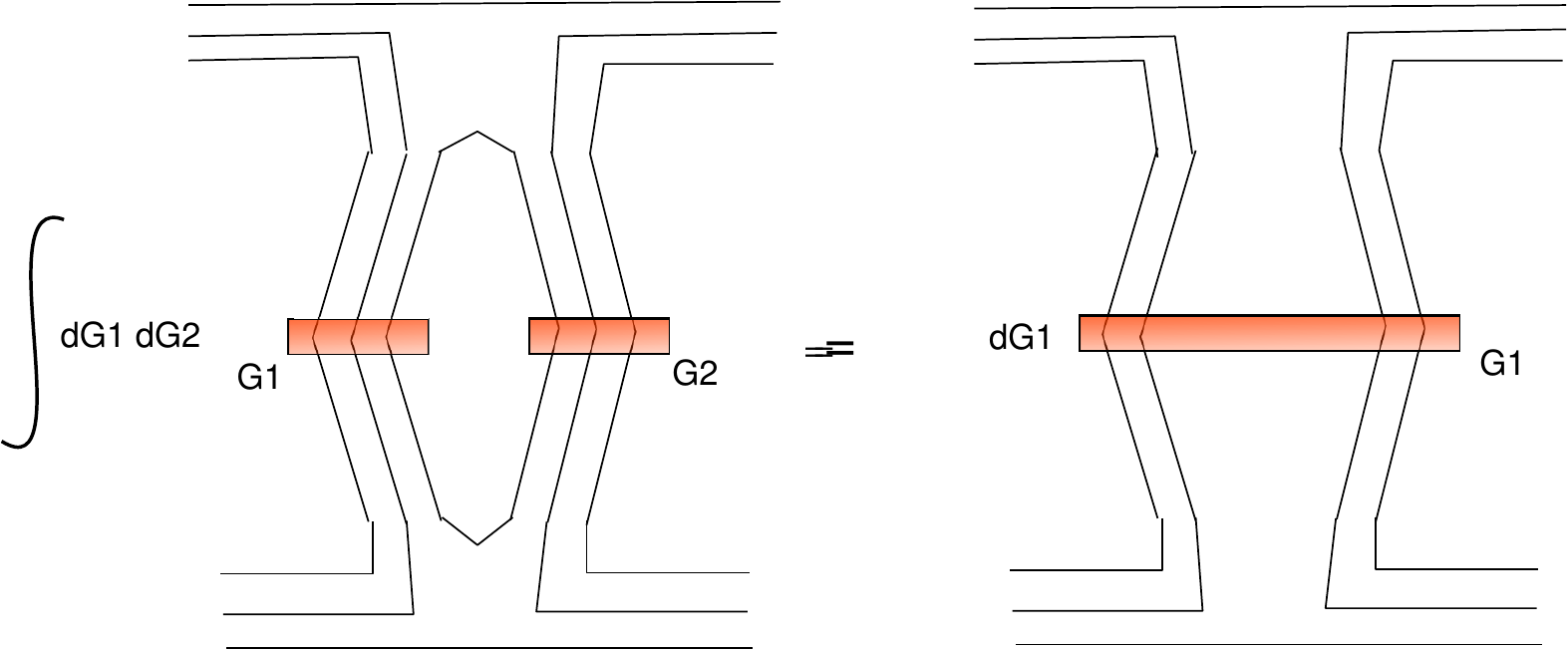}
 \caption{The integration rule for 2-dipoles.}
\label{integral}
\end{figure}

The first connected sum we construct is the one of the sphere, 
which is in graph in Fig. \ref{connsumsph}, which is:
\begin{equation}
S^3\backsimeq[S^3]^n=\underbrace{S^3\ \#\ S^3\ \#\ \cdots\ \#\ S^3}_{n-times}
\label{sphdec} 
\end{equation}
and the amplitude associated with the graph is:
\begin{equation}
 \mathscr A_{2n}(S^{3}) \sim \lambda^{n} \bar \lambda^{n} [\delta^\Lambda(e)]^{n+2}.
\end{equation}
This is topologically an identity, but from the point of view of divergencies is not.  This can be seen from the two graphs in Fig. \ref{connsumsph}.
Evaluating these amplitudes is an exercise, and can be evaluated using the rules in Fig. \ref{integral} and in Fig. \ref{div}. \\\ \\ 

\textbf{Definition} A 4-edge-colored 4-graph is said to be \textit{minimal} for the topological 3-manifold $\mathscr M$ if it is a 3-gem representation
of $\mathscr M$ with the minimal number of vertices.\\\ \\
For instance, the minimal graph for the sphere is made of two vertices, while the minimal graph for the solid 3-Torus is in Fig. \ref{solidtorus}. 

\section{Topological bounds from cuts}
\subsection{Spheres amplitudes}
\begin{figure}[htb]
\center
\includegraphics[scale=0.6]{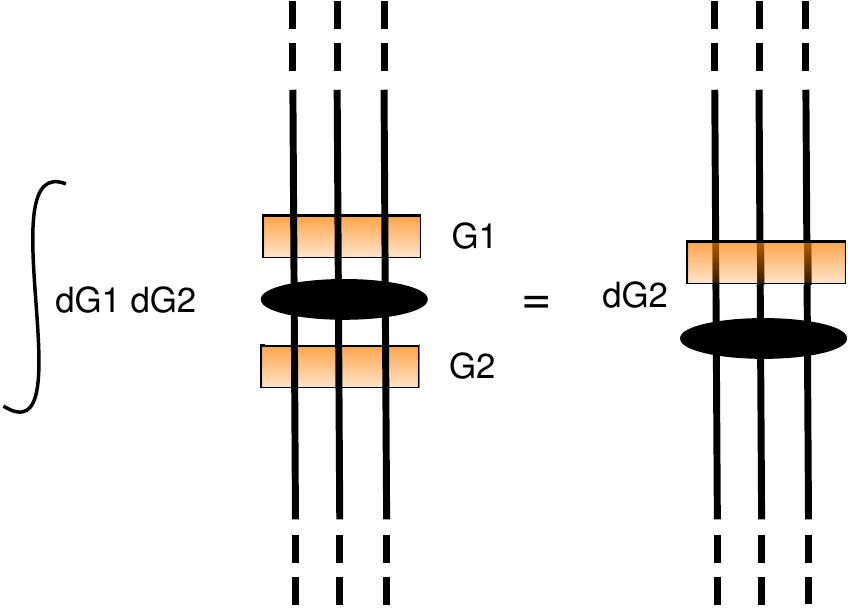}
 \caption{A graphical representation of the identity in eq. \ref{ident1}.}
\label{contr2bubb}
\end{figure}

\begin{figure}[htb]
\center
\includegraphics[scale=0.5]{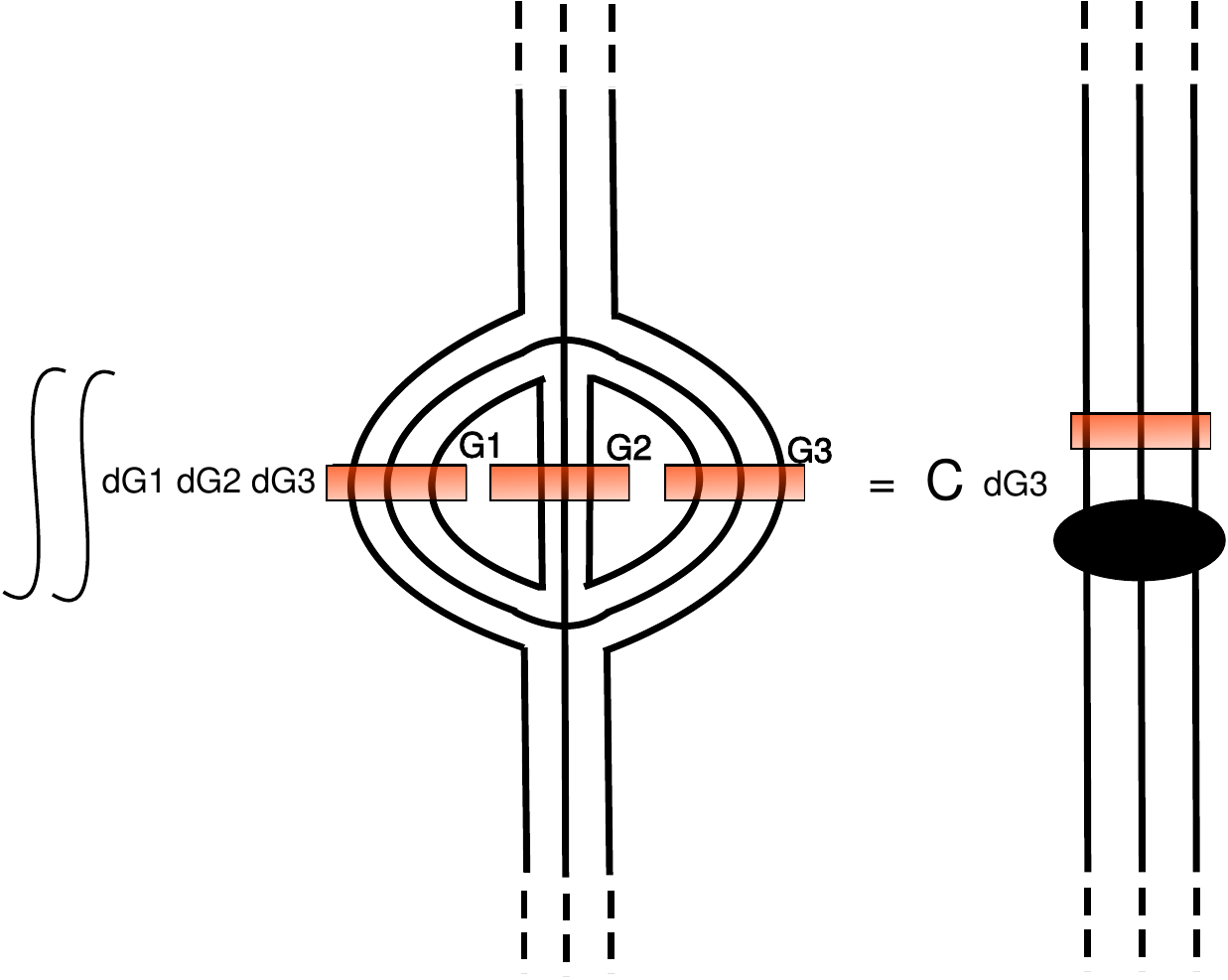}
 \caption{The integration rule for the contraction of a 3-dipole. The constant $C$ is infinite and, regularized, is $\delta^\Lambda(e)$.}
\label{3-dipole}
\end{figure}

As it has been shown recently\cite{1on}, 3-spheres play a special role in 3-dimensional group field theories, role played indeed by 2-spheres in matrix models. It is fact true that 3-balls (in the case of vacuum graphs) dominate the partition function of cGFT in the limit in which the cutoff $\Lambda$ goes to infinity, but not if we keep the cutoff finite other topologies contribute to the partition function.
The result, a posteriori, is not surprising for the following argument. For a generic 4-colored graph $G$ of $2n$-vertices in a $D$-dimensional GFT, the following bound holds on its associated amplitude \cite{verticesbound}:
\begin{equation}
\mathscr A(G)\leq K^{2n} \Lambda^{3(D-1)(D-2) n/2+3(D-1)}
\label{topbound} 
\end{equation}
where $\Lambda$ is the cutoff and $K$ a constant. For $D=3$ this bound reduces to:
\begin{equation}
\mathscr A(G)\leq K^{2n} \Lambda^{3 n/2+6}
\label{topbound3} 
\end{equation}
In particular it is easy to prove that the following upper and lower bounds are valid for the sphere:
\begin{equation}
c_1 \Lambda^6 \leq  \frac{ \mathscr A_{2n}(S^3) } { \lambda^{n} \bar \lambda^n } \leq c_2 \Lambda^{3(n+2)}.
\end{equation}
in a certain sense these bounds are trivial because the lower is (\ref{topbound3}) with no vertices and the upper is exactly the  upper bound for a generic graph.
These bounds are less trivial, however, because for the 3-sphere these are saturated by the two graphs in Fig. \ref{connsumsph}, the upper by the graph above, the lower by the graph below. Here, however, we want to stress that
it is this peculiar feature of divergencies itself which make the analysis of graphs associated with manifolds of higher dimension much more complicated. For this reason, in what follows, we will put forward a procedure in order to factorize diverging terms due to the sphere. We now introduce few identities which will turn useful later. These are in Fig. \ref{contr2bubb} and Fig. \ref{3-dipole}.

We first notice two identities. The first is the one in Fig. \ref{contr2bubb}, which comes from the following identity:
\begin{eqnarray}
&\int \int dg_1 dg_2 \delta(A_1 g_1 g_2^{-1} B_1) \delta(A_2 g_1 g_2^{-1} B_2) \delta(A_3 g_1 g_2^{-1} B_3) =  \nonumber\\ 
& \int dg_1\delta(A_1 g_1 B_1) \delta(A_2 g_1 B_2) \delta(A_3 g_1 B_3) 
\label{ident1}
\end{eqnarray}

\subsection{Cuts, bounds and sphere}
\begin{figure}[htb]
\center
 \includegraphics[scale=0.4]{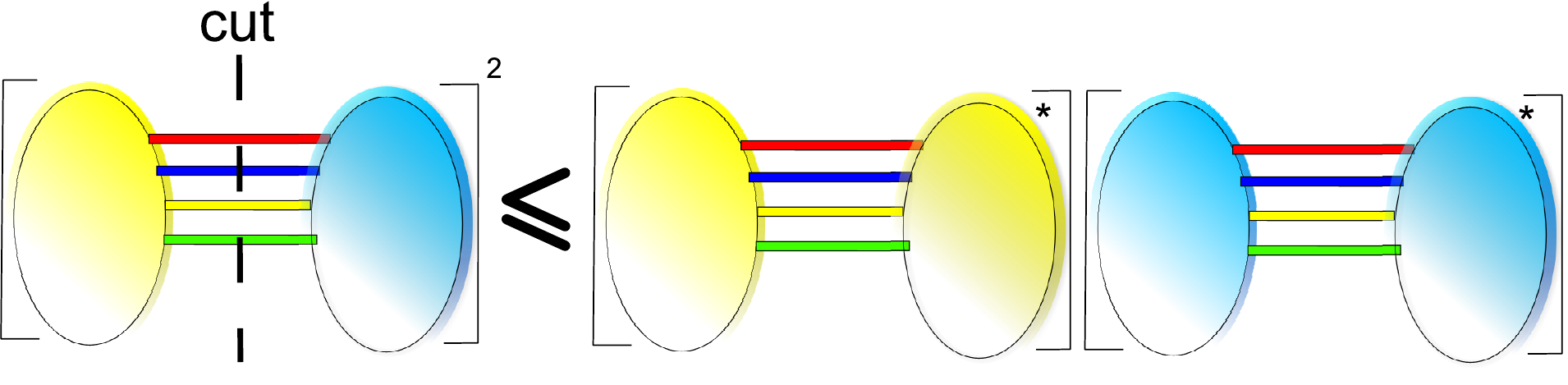}
  \caption{Cut rule for amplitudes. The cut can be done over all possible propagators. Here we focus on the case in which it cuts off only pieces of graphs homeomorphic to the sphere.}
\label{cut}
\end{figure}

Before going into the body of the paper, let us recall the technique of the cuts studied in \cite{verticesbound}.
For simplicity we discuss colored vacuum graphs, but some of the arguments can be generalized to non-colored graphs with open lines.
Let us consider here vacuum graphs of a GFT in $D$ dimension over the group $G$. Consider two graphs $A$ and $B$ which are
connected by a series of propagators $p_1,\cdots,p_n$ with group elements $g_1,\cdots,g_n \in G$. The total amplitude can be written as:
\begin{equation}
\mathscr A=\int \prod_i dg_i A(g_1,\cdots,g_n) B^*(g_1,\cdots,g_n)   
\label{spr}
\end{equation}
where $d g_i$ represents the Haar measure over the group. Since the functions $A,B$ are real, it is easy to see that (\ref{spr}) has all the properties of a distance
and so can be considered as a scalar product between the function $A$ and $B$, $(A,B)$. Since (\ref{spr}) is a scalar product, then it is bounded by the Cauchy-Schwarz inequality:
$$(A,B)^2\leq (A,A)\ (B,B) $$
We will call \emph{cut} over the propagators $p_1,\cdots,p_n$ the bound given by the Cauchy-Schwarz inequality referred to the two parts of the graph $A$ and $B$,
as depicted in Fig. \ref{cut} for the colored model. In the colored Boulatov model the complex structure require the change of the orientation of every vertex that is changing the color
of the vertex (black $\leftrightarrow$ red). The complex amplitude is then glued with conjugate of itself, as the star represent in Fig. \ref{cut}.

Another
\begin{figure}[htb]
\center
 \includegraphics[scale=0.5]{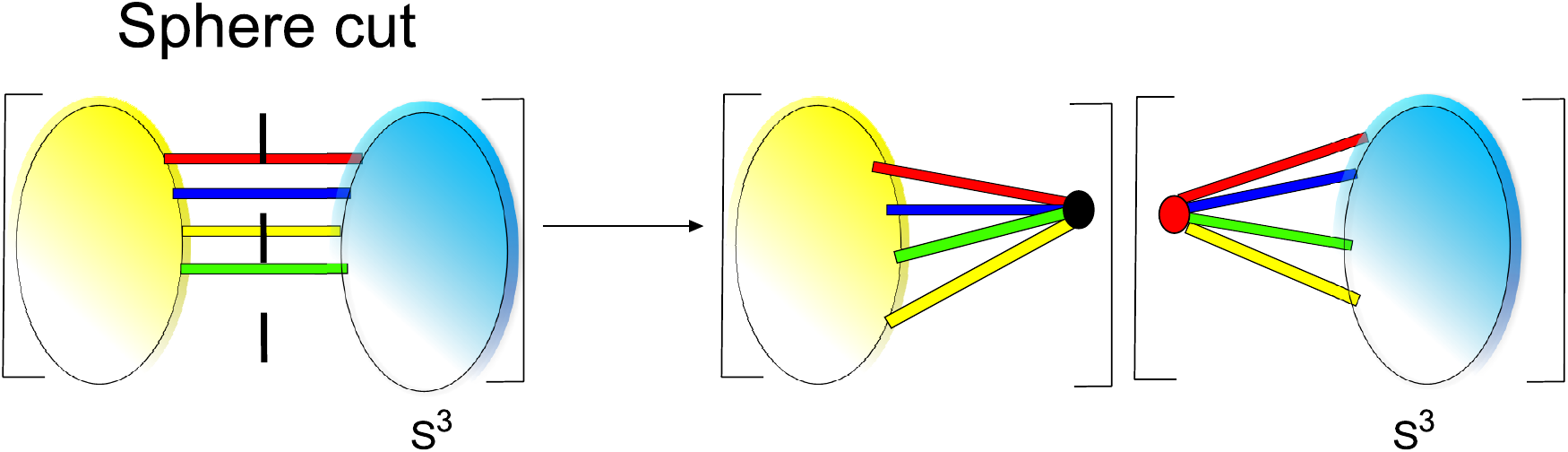}
  \caption{A sphere cut. Cut 4 propagators along the connected sum of the manifold with a piece of graph homeomorphic to a sphere and insert two opposite vertices.}
\label{spherecut}
\end{figure}

Let us now discuss sphere decomposition of graphs and what we will call \textit{sphere cuts}. 
Consider a graph generated by the colored Boulatov model which can be decomposed as in Fig. \ref{spherecut},
 where two part of the graphs are connected by a propagator. We call \textit{topological cut} an insertion of two disconnected 
vertices with different orientation in the propagator where a manifold is connected to the boundary of a 3-Ball. Since the amplitudes
are invariant under color conjugation, we can always redefine the amplitude in such a way that the insertion is as in 
Fig. \ref{spherecut}. We say that we are cutting a sphere if one of the graphs can be reduced through non-degenerate
fusion/contractions of 1- and/or 2- dipoles to a 3-ball (a vertex). This can be extended to more general graphs, where we decompose 
the graph into multi spheres and consider their cuts. In order to see the result of our topological cuts on a concrete example, we apply this to the case of the connected sum of solid torii. 
The reason why we introduce this is because the solid torus is the easier example due to its structure, and to the structure of its core graph. 
\section{Solid Torus decomposition}
\begin{figure}[htb]
\center
\includegraphics[scale=0.3]{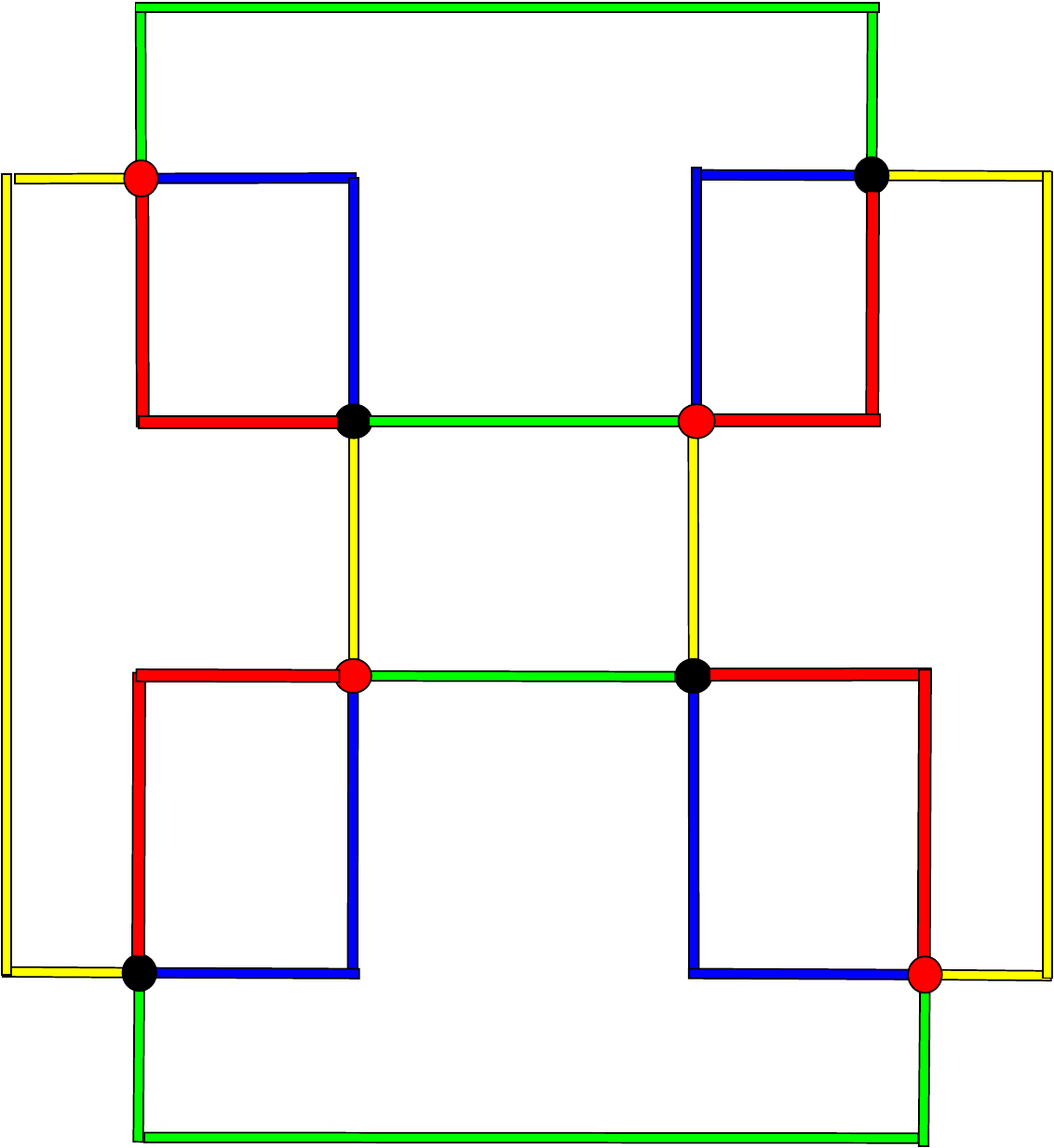}
\caption{The \textit{minimal} graph of the solid torus $S^1 \otimes S^2$.}
\label{solidtorus}
\end{figure}

Consider now the graph in Fig. \ref{solidtorus} for the solid torus \cite{handle1}\cite{handle2}.

\begin{figure}[htb]
\center
\includegraphics[scale=0.5]{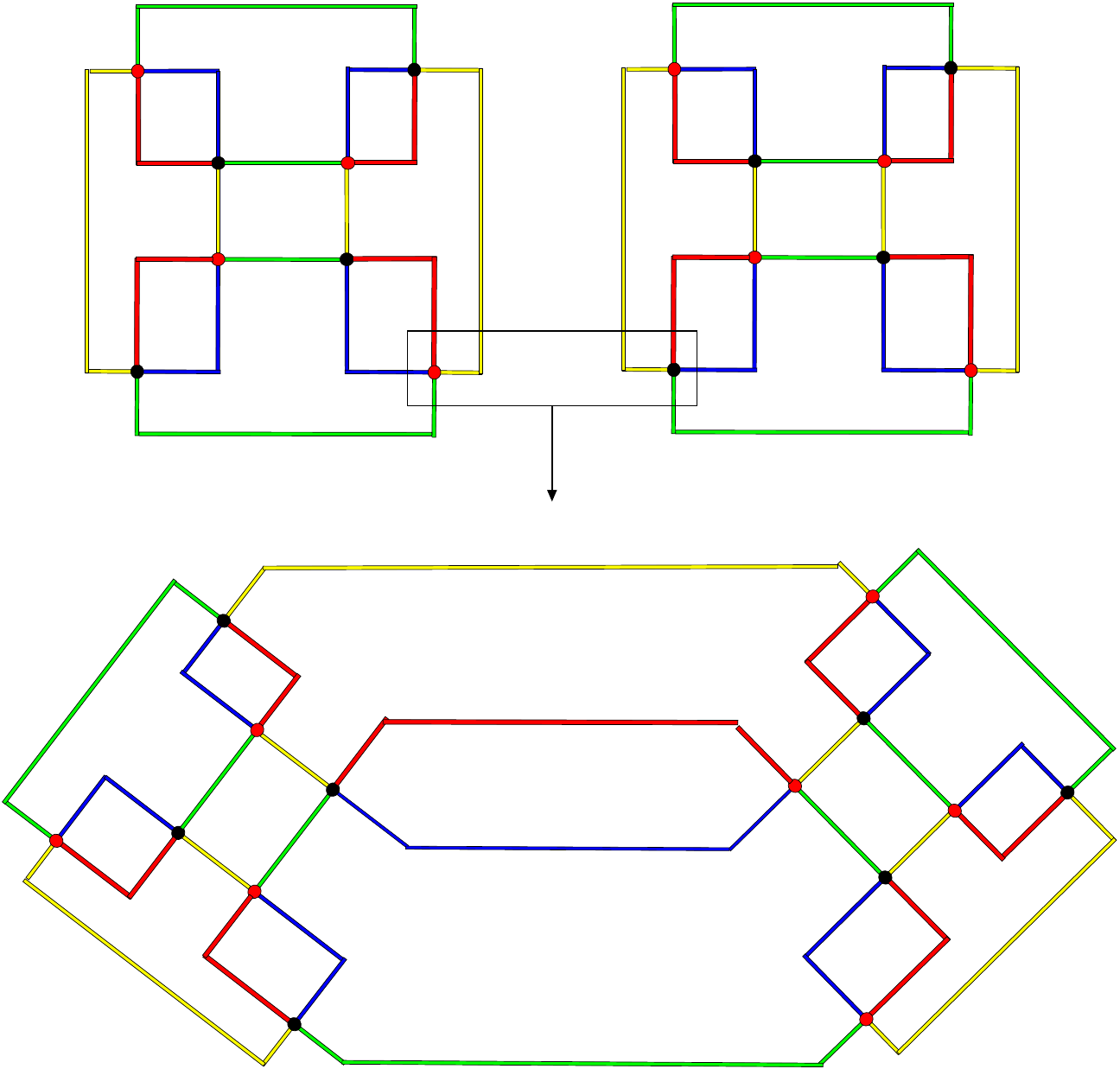}
 \caption{The composition rule for torii graphs: $T_3^2=T_3\ \#\ T_3 $.}
\label{solidtorus_join}
\end{figure}

For a finite number of vertices we can build the following decomposition by mean of a connected sum of solid tori $T_3=S^1 \otimes S^2$:

\begin{equation}
T_3^h=\underbrace{T_3\ \#\ T_3\ \#\ \cdots\ \#\ T_3}_{h-times}.
\label{torusdec} 
\end{equation}

\begin{figure}[htb]
\center
\includegraphics[scale=0.6]{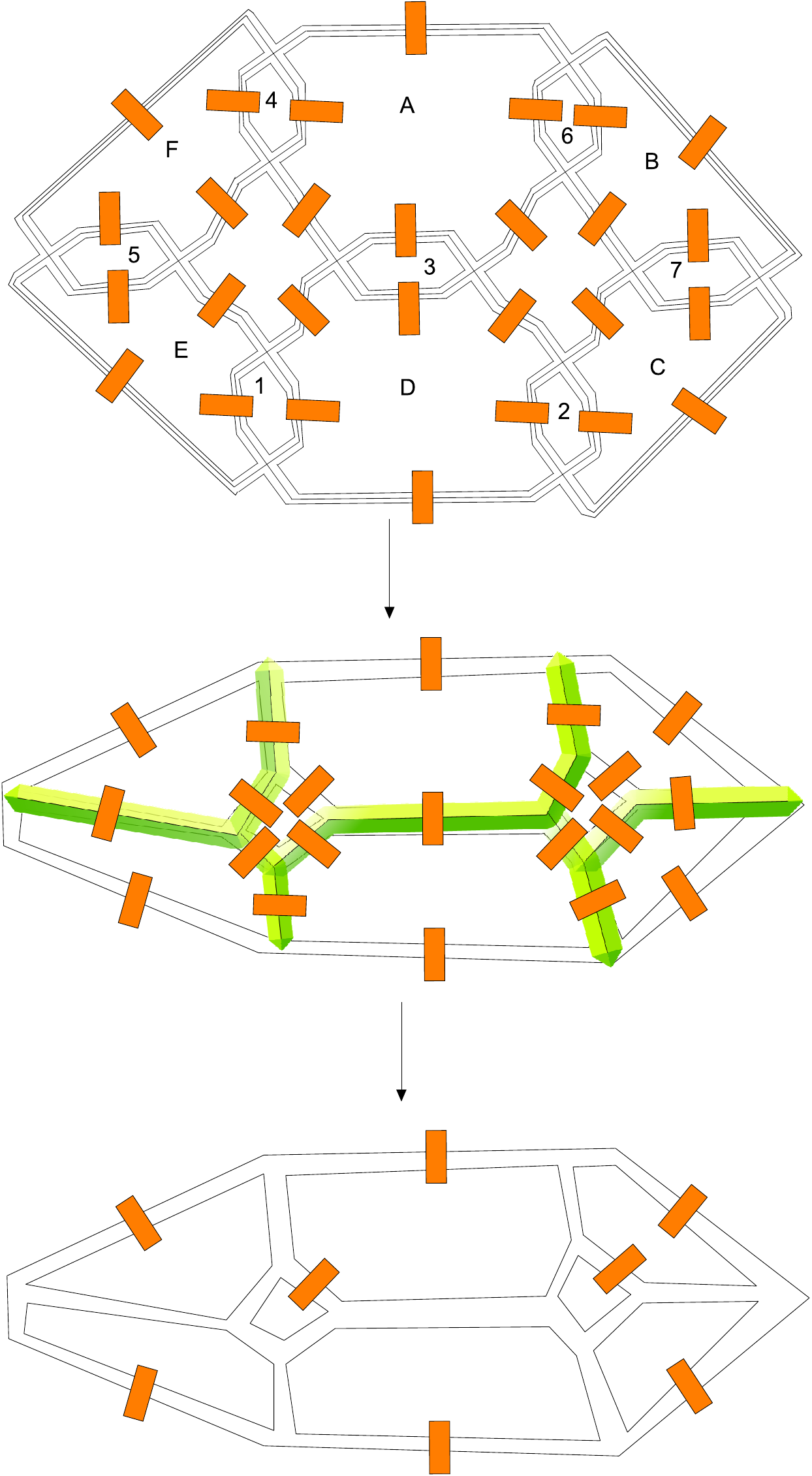}
  \caption{The evaluation of the minimal diagram for $T_3\ \#\ T_3$.}
\label{torus_sempl}
\end{figure}
We now show that the divergence of this manifold with $h-handles$ is related to the number of handles.\\\ \\
\textbf{Lemma} The divergence of the connected sum in (\ref{torusdec}) is for $h\geq1$:
\begin{equation}
\mathscr A \sim \lambda^{1+3h} \bar  \lambda ^{1+3h} [\delta^{\Lambda}(e)]^{2(h+1)}
\label{divtorii}
\end{equation}
\textit{Proof}. Consider the graph in Fig. \ref{solidtorus}. This graph represent a solid torus built with the minimal number of
vertices. By the theorem on the contraction of dipoles introduced antecedently, we can then construct connected sum of solid 3-tori by joining two 
graphs as in Fig. \ref{solidtorus_join}. In order to prove this fact, consider the diagram in Fig. \ref{torus_sempl} and then the generic connected sum is an easier generalization of this one.
In the diagram on top of Fig. \ref{torus_sempl} consider the 2-bubbles labeled with numbers from $1,\cdots,7$. Using the integration rule for 2-dipoles as in \ref{contr2bubb}, the diagram can be transformed with fewer integration over propagators,
as in the second step of Fig. \ref{torus_sempl}. We now notice that there are concentric strands for each of the part of the diagram labeled by capital latin letters, labeled from $A,\cdots,E$. Extract the divergence and put it aside using the integration 
rule for the reduction of concentric strands, as in Fig. \ref{div}. For each latin letter there one power, because there is are 2 concentric strands. Having extracted the
divergence it is easy to notice that what is left is a ribbon graph generated by a 2-dimensional GFT. Pick a maximal tree as in the green shadowed line as in the second step of \ref{torus_sempl} and set the propagators to one. What is left is a 
diagram as in step 3 of Fig. \ref{torus_sempl}. It is easy to see that this diagram is finite. All the steps, also the use of the maximal tree, can be generalized to several connected sums. It is then also easy to see, at this point, that the number of divergencies is related to the parts of
the graph labeled to latin letters and that can be counted easily for generic connected sums. From here it follows Eq. (\ref{divtorii}).\\\ \\

\begin{figure}[htb]
\center
\includegraphics[scale=0.5]{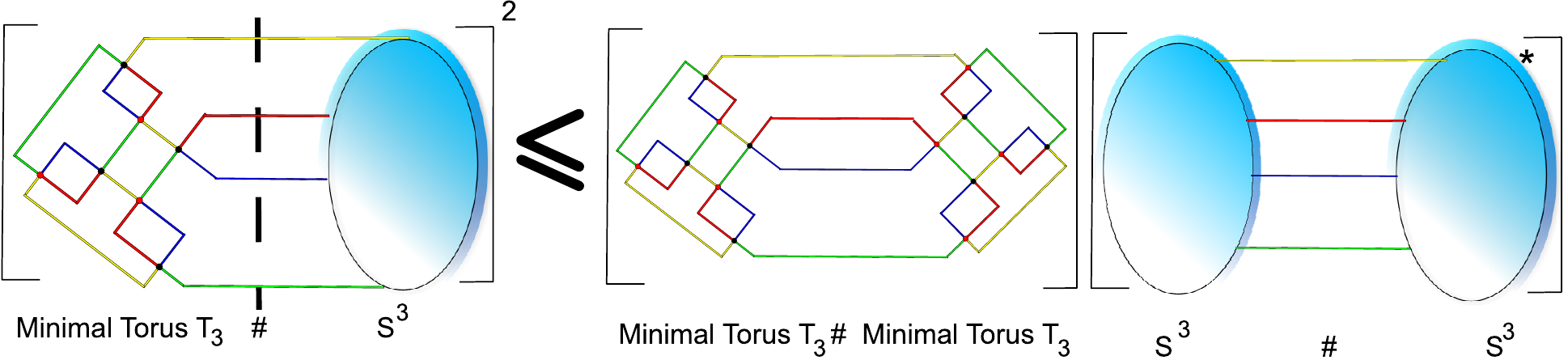}
 \caption{The cut applied on the connected sum of a minimal torus and a sphere. This procedure put the uncertainty of the divergence of the graph on the graph of the sphere, which is easier to study.}
\label{cuttorus}
\end{figure}
\begin{figure}
\includegraphics[scale=0.5]{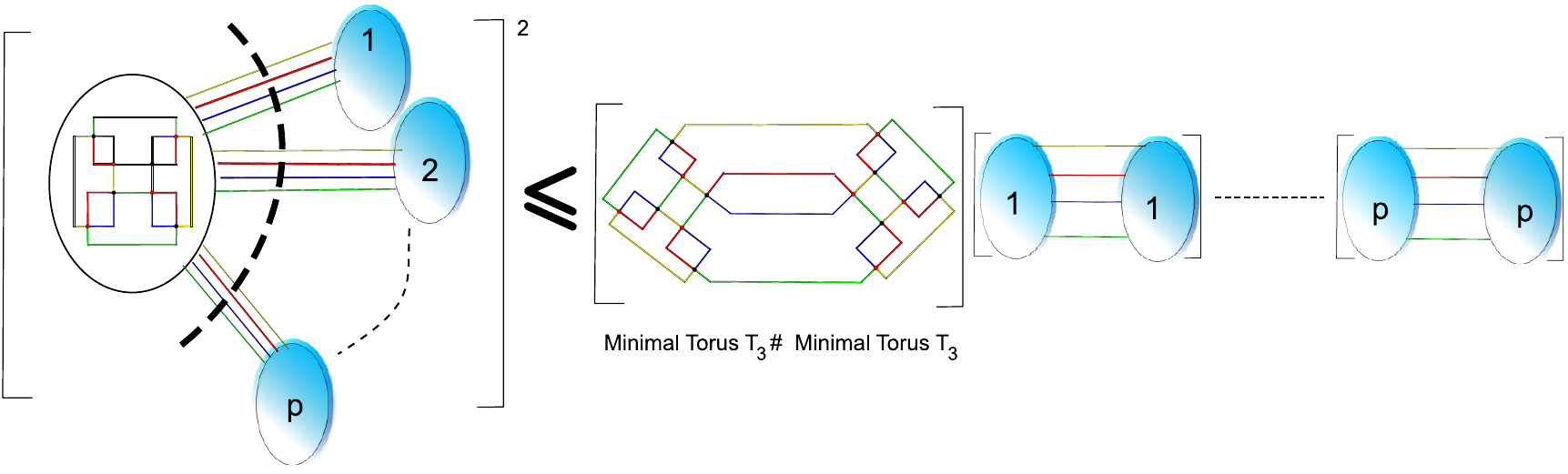}
\label{multcut}
\caption{Multiple cuts to get the terms homeomorphic to $S^3$ cutted off. The result is the same as the one with a single $S^3$ cut off.}
\end{figure}

\subsection{Evaluating the bounds}
\begin{figure}[htb]
\center
\includegraphics[scale=0.5]{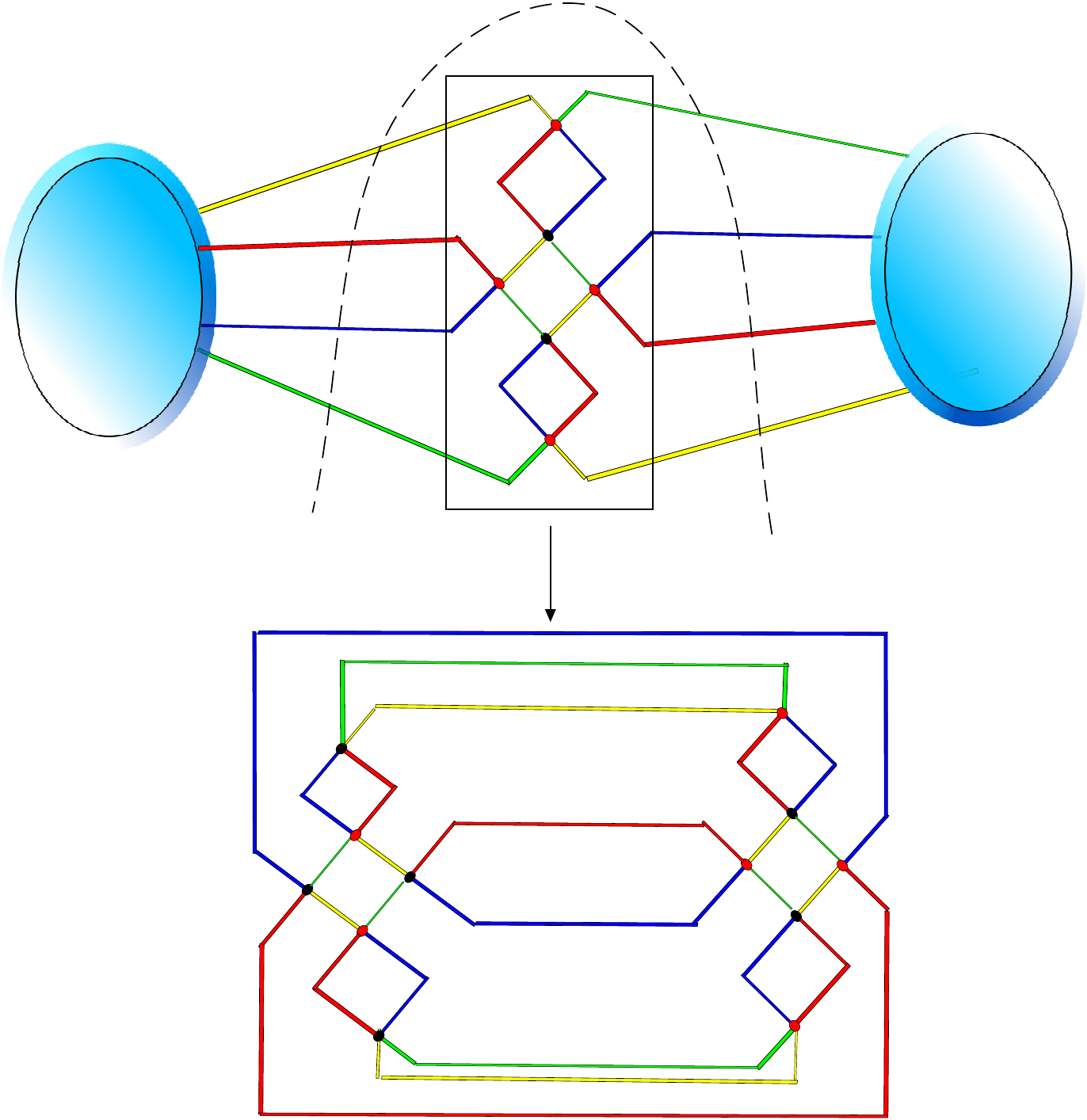}
 \caption{The multiple cut for $T_3\ \#\ M_1\ \#\ M_2$, where $M_2$ and $M_1$ are other two manifolds.}
\label{multcutright}
\end{figure}

We can now make a bound on generic genus-$h$ manifolds as follows. We first recall the following norm as a cut over the propagators of a Feynman graph.
Consider a graph split into two graphs by a series of cuts, as in Fig. \ref{cut}. Then the amplitude square is bounded by the norms of the left
and right graph contracted with themselves by converting a red vertex to a black vertex and viceversa. We can now use this trick in a clever way. 
It is well known in topology that every manifolds is homotopic to itself modulo cutting off $S^3$. We can then think of any manifold, in a simplicial setting
in which this manifold is constructed by $2k$ vertices, as a \textit{minimal} one, made of $2j$ vertices and then a connected sum of 3-spheres by a connected
sum as explained in this section, and leaving $2j-2k$ vertices to this part. In the simplest case in which the simplicial complex is a minimal one and the connected sum
of a sphere we can consider the cut along the two vertices which make the connected sum, as in Fig. \ref{cuttorus}.
We now evaluate a bound on the divergence of graph representing $T^h_r$ using \ref{cut}. It is well known in topology that every manifolds is homotopic to itself modulo cutting off $S^3$. 
Following our the previous discussion, the bounds comes from taking cuts\cite{bosoniccgft} over edges on the connected sum of the minimal torus. 
\begin{equation}
\mathscr A ^2_{2k}(T_3) \leq \mathscr A_{14}(T^3\ \#\ T^3) \mathscr A_{2(k-7)} (S^3)
\end{equation}
where, however, now the divergence of the graph of $T^3\ \#\ T^3$ is minimal and has been evaluated in (\ref{divtorii}) for generic genus. Thus the bound is on the divergence of the sphere, for which we can use the bounds introduced before. This kind of \textit{homeomorphic} bounds can be generalized to other types of homeomorphic manifolds as, for instance, connected sums of torii, as we will do later in the paper. In this case the bound takes the following form:
\begin{eqnarray}
& \mathscr A ^2_{2k}(\underbrace{T_3\ \#\ T_3 \#\ \cdots\ \#\ T_3}_{n-times}) \leq \nonumber \\
& \leq\mathscr A_{6n+2}(\underbrace{T_3\ \#\ T_3 \#\ \cdots\ \#\ T_3}_{2n-times}) \mathscr A_{2k-6n-2} (S^3)
\label{boundtori}
\end{eqnarray}
where you cut only one sphere. In this case each connected sum of torii with $h$ handles is bounded by a canonical $2h$ handles, which is evaluated in general,
and multiple amplitudes on the spheres, depending whether there are one or more cuts, as in Fig. \ref{cuttorus}. This identity is related to the well known fact $\mathscr M \backsimeq \mathscr M\ \#\ S^3$. This formula shows that the ``uncertainty'' of the divergencies of a generic triangulation of a 3-manifold
can be split into a canonical decomposition thanks to the theorem on the contraction of 1-dipoles and put in relation with solid tori, and put the uncertainty into the different graph homeomorphic to the spheres. There is, however, a subtlety due to the fact that in general a manifold can be seen as many connected 
sums of the same manifold with spheres. In this case a feasible cut should on multiple propagators connecting spheres, as in Fig. \ref{multcutright}. However, it is easy to understand that each time you make a \textit{single} cut between an h-handles diagram and a sphere, the diagram for the handles part is a connected sum of the same two h-handles and thus
becomes a 2h handles diagram. However, when there are multiple cuts the same diagram is connected multiple times with itself. It is easy to understand that making a unique cut along all the pieces homeomorphic to spheres can glue torii with themselves more than once, 
making them noncanonical, and thus difficult to recognize the handle decomposition. In principle this should be performed one cut at time. However, performing one cut at time double the number of pieces homeomorphic to the sphere. A possible trick would be to make this an infinite number of times:
\begin{eqnarray}
&\mathscr A^2 (\underbrace{T_3\ \#\ \cdots\ \#\ T_3}_{k-times}\ \#\ \underbrace{S_3\ \#\ \cdots \ \#\ S_3}_{p-vertices}) \leq \nonumber \\
& \leq \lim_{n\rightarrow \infty} \{ \mathscr A (\underbrace{T_3\ \#\ \cdots\ \#\ T_3}_{2nk-times}\ \#\ \underbrace{S_3 \ \#\ \cdots\ \#\ S_3}_{the\ rest})\}^{\frac{1}{n}}  \cdot \nonumber \\
& \cdot \sqrt{A_{i_n}(S_3)\sqrt{ \cdots \sqrt{ \mathscr A_{i_2}(S_3) \sqrt{\mathscr A_{i_1}(S_3)}}}} \mathscr A_{i_0}(S_3)
\label{complicatedformula}
\end{eqnarray}

\begin{figure}[htb]
\center
\includegraphics[scale=0.5]{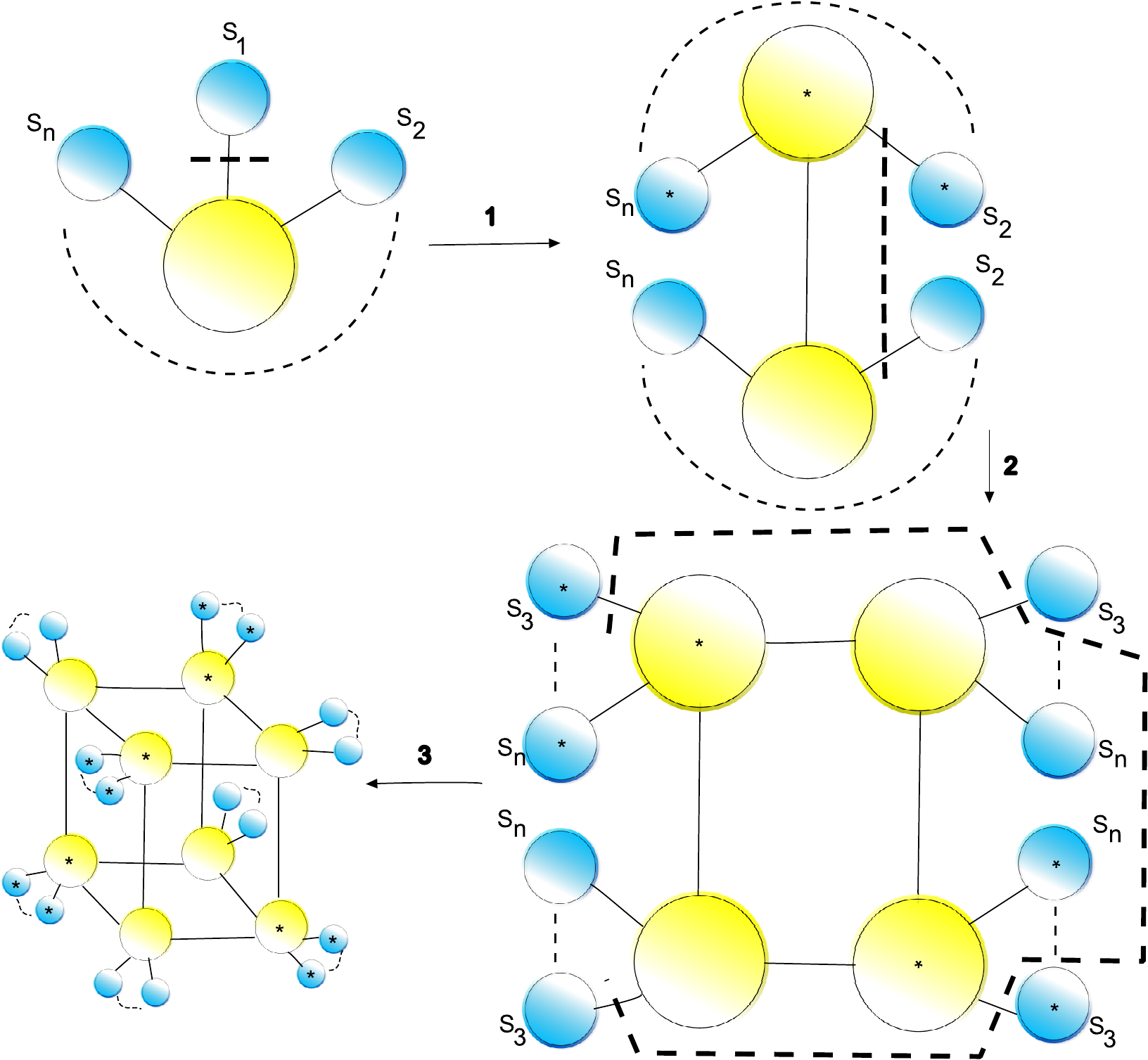}
\caption{The first three cuts of the sequence for a connected with $n$ different spheres, $S_i$.}
\label{multcuts}
\end{figure}

Now you would expect that in the $n\rightarrow\infty$ limit $$\mathscr A (\underbrace{T_3\ \#\ \cdots\ \#\ T_3}_{2nk-times}\ \#\ \underbrace{S_3 \ \#\ \cdots\ \#\ S_3}_{the\ rest}),$$ 
becomes a canonical connected sum of torii, which we know how to evaluate. 
However, a careful thought will convince that this cannot happens. The reason is that while the number of cuts increases linearly the number of spheres increases as $2^n$. Thus in order to extract the divergencies due to the sphere in the bound, a neat sequences of cuts
have to be made. In the following we will, for simplicity, represent a propagator as a solid line, cuts as bold dashed lines, as in Fig. \ref{multcuts}.

\begin{figure}[htb]
\center
\includegraphics[scale=0.5]{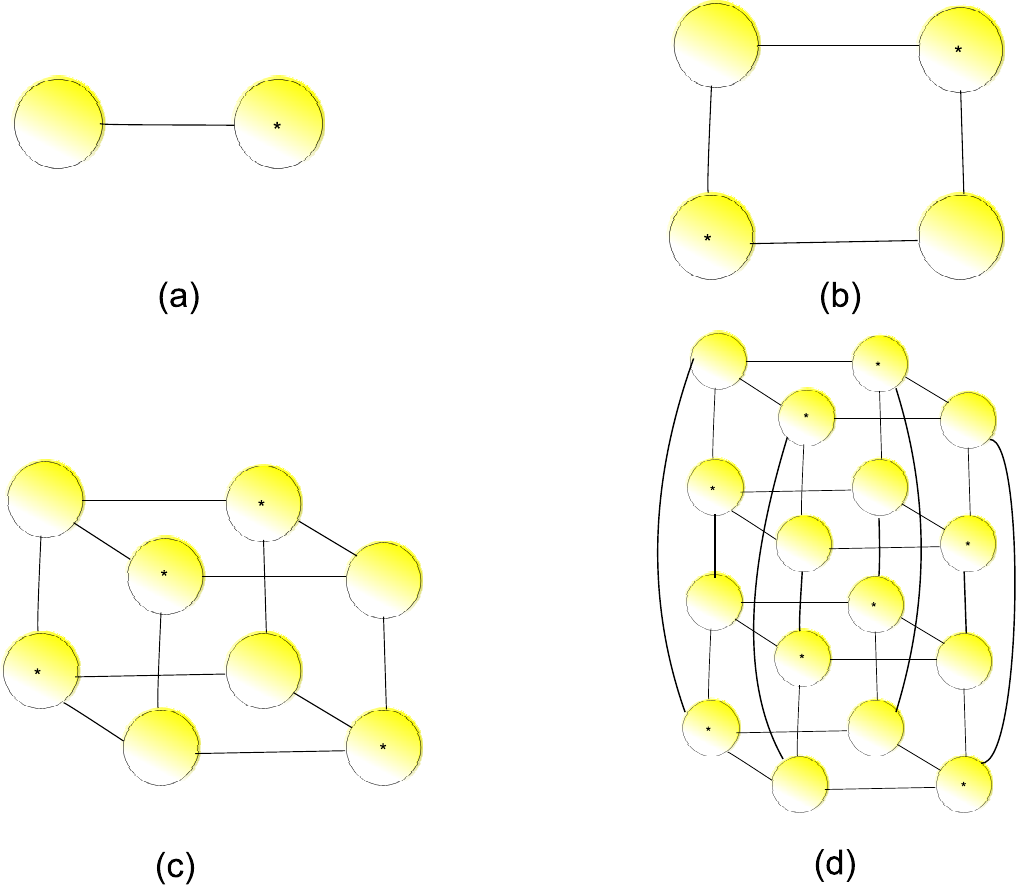}
 \caption{In (a)-(d), the graphs $Q^n$ generated for $n=1,\cdots,4$.}
\label{graphs4}
\end{figure}

As we have shown before, an infinite number of cuts limit is not feasible. Thus we have to choose the 
sequence of cuts in a clever way. In particular, note that when you perform a cut the number of spheres double.
The trick is to double also the number of spheres cut at each step. 
For instance as in Fig. \ref{multcuts}, the first cut is on one sphere, the second cut is on the two $S_2$ spheres, the third on the four
$S_3$ spheres and so on. In this way, the number of cuts is $n$ before we exhaust the number of spheres. 
Each $n$-th cut separates $2^n$ spheres. Carrying all the cuts until the ends give the following bound:
\begin{equation}
\mathscr A (G\ \# S^3_1\ \#\ \cdots\ \# S^3_n)\leq \ (\mathscr A(\mathscr Q^n))^{\frac{1}{2n}} \prod_{i=1}^n \sqrt{A(S_i\ \#\ S_i)} 
\label{finalf}
\end{equation}
Now $\mathscr Q^n$ is a very particular graph which depends on the number of spheres but has a typical form for
$n>2$, and which we will now explain. The case $n=1$ is special. To give a graphical representation, if the amplitude was a single vertex, the resulting amplitude at the cut
$(n-1)th$ cut would be a piece of $(n-1)$-dimensional cubic lattice where at each point the $S_n$ sphere is attached. The case $n=1$ is trivial. In Fig. \ref{graphs4} the graphs generated by the sequence of cuts in the
cases $n=1,\cdots,4$. Notice that these cuts are well behaved from the topological point of view. In fact, when we do a cut on a sphere, for instance in the $n=1$ case, what we are doing is the following:
take a manifold and remove a ball $B^3$ from it, then take the same manifold, reverse the orientation, carve a ball $B^3$ and identify the boundaries between the two. This
is the same as ordinary surgery in topology. Let us define now the following surgery operation $\#^n$: carve $n$ balls out of a manifolds, copy the manifold with opposite orientation
and identify the boundary of these balls pairwise. We now have all the elements to give a proper definition for the quantity $\mathscr Q^n$. Let $\mathscr G$ be the topological manifold associated with $G$.  Then the resulting $\mathscr Q^n$ manifold is contructed by recursively cutting and pasting as follows: 
\begin{eqnarray}
& \mathscr Q^1=\mathscr G\ \#\ \mathscr G \nonumber \\
& \mathscr Q^2=(\mathscr G\ \#\ \mathscr G)\ \#^2\ (\mathscr G\ \#\ \mathscr G)\nonumber \\
\end{eqnarray}
In general, this is defined for a generic $n$ as:
\begin{equation}
\mathscr Q^n= \mathscr Q^{n-1}\ \#^n\ \mathscr Q^{n-1}
\end{equation}

Thus, eqn. (\ref{finalf}) gives a compact bound, trivializing the contribution of the spheres if these are strategically separated by means of cuts.  For the case of solid torii, in particular, the divergence can be evaluated and related to the number 
of handles, which behave nicely under the surgery explained. 
In principle this bound, given the smallest representation of
a topological manifold in term of simplices, can be explicitly evaluated. On the other hand, the evaluation of the graph $\mathscr Q^n$ can be quite cumbersome. Another possible cut, which is quite economical, is given by
the single cut of all the spheres together. This cut, once performed, leaves us with what we will care a "core graph". For the $n=8$ case is for this cut the simplest to evaluate for the solid torus, while it is $n=v_{vert}$ case for a generic graph with $v_{vert}$ vertices. Using the integration rule for concentric strands the resulting graph becomes counting the number of circles in the graph. In this case, in fact, there is a $\delta^{\Lambda}(e)^2$, because a propagator
is made of 3 internal strands.
For the solid torus $n$ can be as big as $8$, which would be the most general case to evaluate. To give an example of how to evaluate this bound, we evaluate the cases
$n=1,\cdots,4$ for the solid torus, showing how it works for $n=2$ and give the final results for $n=3,4$.

\begin{figure}[htb]
\center
 \includegraphics[scale=0.5]{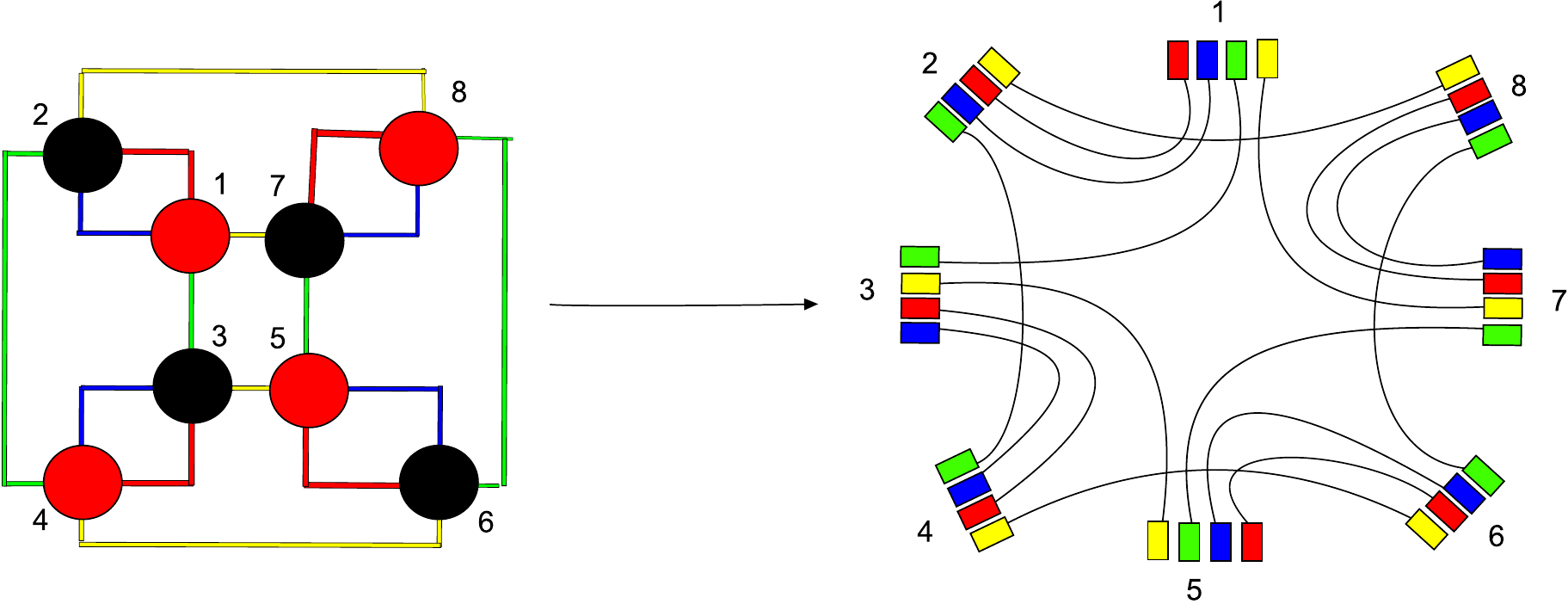}
  \caption{The core graph associated to the $n=8$ minimal representation of the solid torus. To each sphere replace the 4 colored propagators associated with the sphere of the cut. Each line line in the graphical is associated with a stranded line.}
\label{nequal8}
\end{figure}
\begin{figure}[htb]
\center
\includegraphics[scale=0.5]{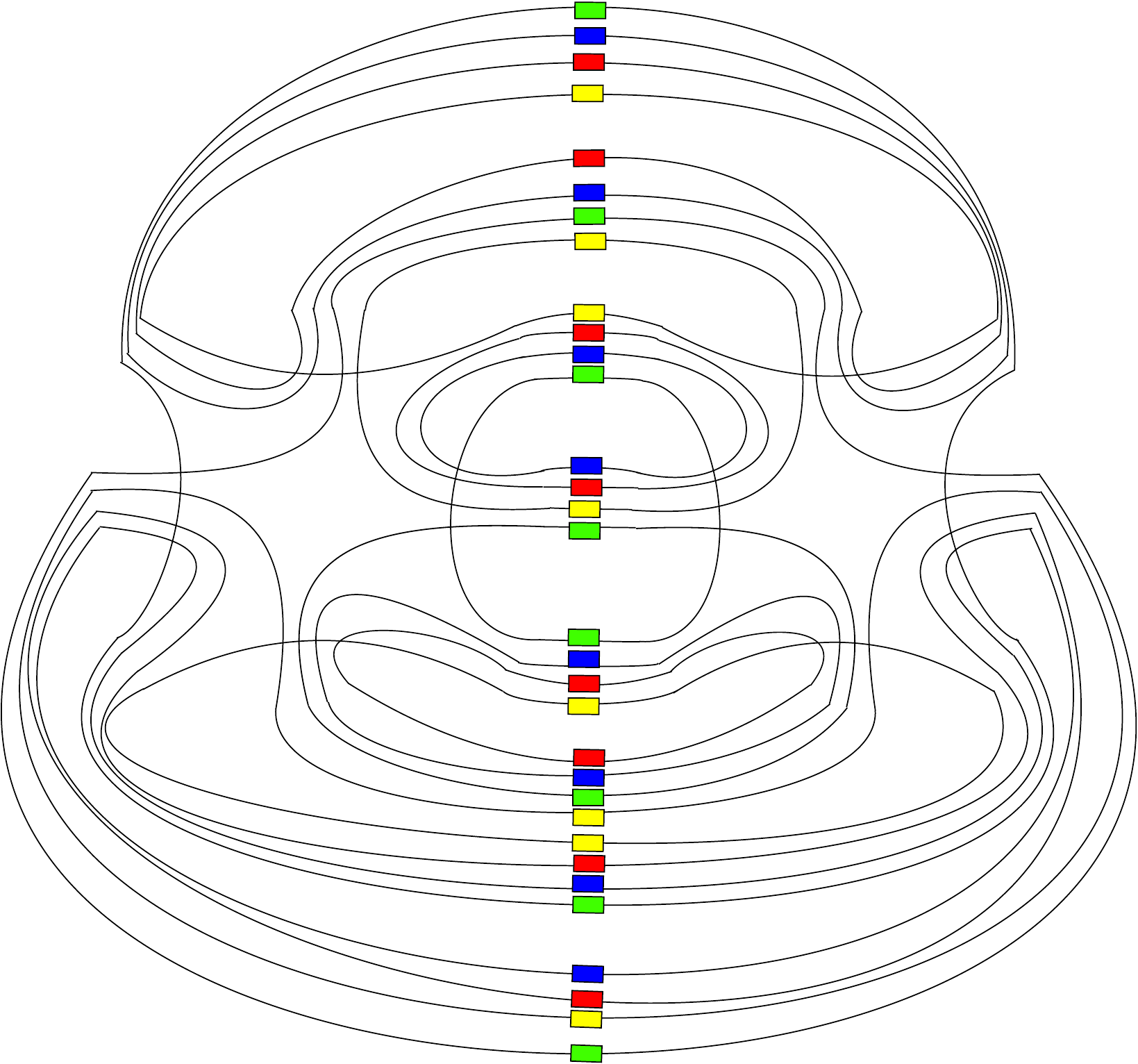}
  \caption{The graph associated with the $n=8$ core graph of the solid torus associated to the cut of all the balls together. The divergence of this graph can be easily evaluated to be $(\delta^\Lambda(e))^{16}$.}
\label{compec}
\end{figure}

\subsection{A \textit{n\"{a}ive} large \textbf{N} limit for Solid Torii}
We now have the possibility, using this exact result for canonical graphs and the motivated conjecture on the maximum divergence of the sphere, to perform a 't Hooft limit in the couplings. Let us consider now the renormalized couplings:
$$\lambda \rightarrow \frac{\lambda_0}{\delta^\Lambda(e)};\ \  \bar \lambda \rightarrow \frac{\bar \lambda_0}{\delta^\Lambda(e)}$$
Using the upper bound, we can see that as:
$$sup\ \mathscr A_{2k}(S^3)\sim  [\delta^{\Lambda}(e)]^{1-n}$$
and:
$$\mathscr A_{2k}(T_3^h)\sim [\delta^{\Lambda}(e)]^{-4 h}$$
It is easy to see that the first order of the sphere does not diverge, while the higher the number of handles, the higher the graph is suppressed in the weak coupling limit. This is compatible with the result obtained in \cite{1on}.

It is, however, easy to understand that these are valid for \textit{minimal} graphs for solid torii. Let us then use the bound given by (\ref{boundtori}) given by cutting only one sphere of $2n$ vertices. In this case:
$$sup\ \ \mathscr A_{2k}(T_3^h)\sim [\delta^{\Lambda}(e)]^{-4 h+n+1}$$
thus there is a competition between the number of handles of the canonical graph and the vertices associated to the sphere which has been cut.

\section{Conclusion}

Group field theories provide one of the most
promising framework for a background-free theory of quantum gravity in which one sums
both over topologies and geometries. The important feature of the model rests on the structure of the divergencies,
related to the structure of the Feynman graphs. It has been shown, under a proper regularization of the integrals preserving
the group structure underlying the theory, that graphs topologically equivalent to sphere dominate the partition function when
the only cutoff is sent to infinity in the renormalized couplings. If the coupling is kept finite, however, other topologies
contribute to the partition function, and indeed it has been shown that they are combinatorially favored in the expansion to the spherical topology.
In this paper, we focused on the topology related to the dual of the Feynman graphs in the perturbative
expansion of the partition function and introduced a formalism of cuts in order to separate divergencies related to different topologies. 
We believe this to be a novel technique aimed at analyzing and factorizing the contribution to the divergence degree due to the sphere.  
These techniques relied heavily on the graph coloring, thus valid only in a colored group field theory, and are aimed at constructing amplitude 
bounds using propagator cuts. We showed that cutting along propagators, the dual can be interpreted as points in which a connected sum of 
two manifolds is performed and can factorize the divergencies of the two topologies.  In particular, if the cut is performed where a sphere is 
connected, then a specific bound can be taken as spheres are very well studied in group field theory. Color is a fundamental ingredient 
in this paper as the connected sum can be defined only on the base of dipole contractions.
After having introduced the technique, we applied it to the connected sum of solid torii as an example.

Although several powerful and precise results on the divergence of generic Feynman diagrams already exist \cite{smerlakbom}, in relation to both the group and the structure of the graph $\Gamma$,
it is clear that the divergences of a Feynman diagram are not related only to the topology of the bubbles. Indeed, the divergencies are intertwined with the topology of the dual of the graph $\Gamma$ as well. 
In this respect, we believe that such formalism introduced in the present paper could contribute to the understanding of the contributions related to the topology of the graph. The sphere cuts introduced here are
indeed made in order to introduce in a clear way topological surgery into the matter of calculating divergence bounds of a particular graph.

The limitation of such approach is its lack of generality in providing cut sequences for generic topologies of the dual graph. However, the sphere cut formalism might be useful in the study of graphs of specific topology.
It has been for long stressed that the partition function of quantum gravitational theory has to have a sum over all the possible topology. In this case in particular, our approach might be helpful in understanding the different behaviors, in terms of divergencies,
of two different topologies.


\begin{backmatter}

\section*{Acknowledgements}
We would like to thank Daniele Oriti, Sylvain Carrozza, Joseph Ben-Geloun and Dan Lynch for several discussions. Moreover, we would like to thank the anonymous referees for the valuable comments.
This research has been mostly conducted while visiting the Albert Einstein Institute (MPI) in Berlin, where also financial support has been given.


\bibliographystyle{bmc-mathphys} 
\bibliography{bmc_article}      

\begin{thebibliography}{34}
\ifx \bisbn   \undefined \def \bisbn  #1{ISBN #1}\fi
\ifx \binits  \undefined \def \binits#1{#1}\fi
\ifx \bauthor  \undefined \def \bauthor#1{#1}\fi
\ifx \batitle  \undefined \def \batitle#1{#1}\fi
\ifx \bjtitle  \undefined \def \bjtitle#1{#1}\fi
\ifx \bvolume  \undefined \def \bvolume#1{\textbf{#1}}\fi
\ifx \byear  \undefined \def \byear#1{#1}\fi
\ifx \bissue  \undefined \def \bissue#1{#1}\fi
\ifx \bfpage  \undefined \def \bfpage#1{#1}\fi
\ifx \blpage  \undefined \def \blpage #1{#1}\fi
\ifx \burl  \undefined \def \burl#1{\textsf{#1}}\fi
\ifx \doiurl  \undefined \def \doiurl#1{\textsf{#1}}\fi
\ifx \betal  \undefined \def \betal{\textit{et al.}}\fi
\ifx \binstitute  \undefined \def \binstitute#1{#1}\fi
\ifx \binstitutionaled  \undefined \def \binstitutionaled#1{#1}\fi
\ifx \bctitle  \undefined \def \bctitle#1{#1}\fi
\ifx \beditor  \undefined \def \beditor#1{#1}\fi
\ifx \bpublisher  \undefined \def \bpublisher#1{#1}\fi
\ifx \bbtitle  \undefined \def \bbtitle#1{#1}\fi
\ifx \bedition  \undefined \def \bedition#1{#1}\fi
\ifx \bseriesno  \undefined \def \bseriesno#1{#1}\fi
\ifx \blocation  \undefined \def \blocation#1{#1}\fi
\ifx \bsertitle  \undefined \def \bsertitle#1{#1}\fi
\ifx \bsnm \undefined \def \bsnm#1{#1}\fi
\ifx \bsuffix \undefined \def \bsuffix#1{#1}\fi
\ifx \bparticle \undefined \def \bparticle#1{#1}\fi
\ifx \barticle \undefined \def \barticle#1{#1}\fi
\ifx \bconfdate \undefined \def \bconfdate #1{#1}\fi
\ifx \botherref \undefined \def \botherref #1{#1}\fi
\ifx \url \undefined \def \url#1{\textsf{#1}}\fi
\ifx \bchapter \undefined \def \bchapter#1{#1}\fi
\ifx \bbook \undefined \def \bbook#1{#1}\fi
\ifx \bcomment \undefined \def \bcomment#1{#1}\fi
\ifx \oauthor \undefined \def \oauthor#1{#1}\fi
\ifx \citeauthoryear \undefined \def \citeauthoryear#1{#1}\fi
\ifx \endbibitem  \undefined \def \endbibitem {}\fi
\ifx \bconflocation  \undefined \def \bconflocation#1{#1}\fi
\ifx \arxivurl  \undefined \def \arxivurl#1{\textsf{#1}}\fi
\csname PreBibitemsHook\endcsname

\bibitem{daniele}
\begin{bchapter}
\bauthor{\bsnm{D.Oriti}}:
\bctitle{The microscopic dynamics of quantum space as a colored group field
  theory}.
In: \beditor{\bsnm{G.~Ellis}, \binits{A.W.} \bsuffix{J.~Murugan}} (ed.)
\bbtitle{Foundations of Space and Time: Reflections on Quantum Gravity}.
\bpublisher{Cambridge University Press},
\blocation{Cambridge}
(\byear{2011})
\end{bchapter}
\endbibitem

\bibitem{razjim}
\begin{botherref}
\oauthor{\bsnm{Gurau}, \binits{R.}},
\oauthor{\bsnm{Ryan}, \binits{J.P.}}:
Colored tensor models - a review.
SIGMA 8, 020; arXiv:1109.4812
(2012)
\end{botherref}
\endbibitem

\bibitem{quantugeom2}
\begin{botherref}
\oauthor{\bsnm{Oriti}, \binits{D.}}:
The group field theory approach to quantum gravity: some recent results.
arXiv:0912.2441
(2009)
\end{botherref}
\endbibitem

\bibitem{sf}
\begin{botherref}
\oauthor{\bsnm{Perez}, \binits{A.}}:
The spin foam approach to quantum gravity.
Living Rev. Relativity 16 , 3; arXiv:1205.2019
(2013)
\end{botherref}
\endbibitem

\bibitem{Brezin:1977sv}
\begin{botherref}
\oauthor{\bsnm{Brezin}, \binits{E.}},
\oauthor{\bsnm{Itzykson}, \binits{C.}},
\oauthor{\bsnm{Parisi}, \binits{G.}},
\oauthor{\bsnm{Zuber}, \binits{J.B.}}:
Planar diagrams.
Commun.\ Math.\ Phys.\ {\bf 59}, 35
(1978)
\end{botherref}
\endbibitem

\bibitem{mm}
\begin{botherref}
\oauthor{\bsnm{David}, \binits{F.}}:
A model of random surfaces with nontrivial critical behavior.
Nucl.\ Phys.\ B {\bf 257}, 543
(1986)
\end{botherref}
\endbibitem

\bibitem{sing}
\begin{botherref}
\oauthor{\bsnm{Gurau}, \binits{R.}}:
Lost in translation: Topological singularities in group field theory.
Class.\ Quant.\ Grav.\ {\bf 27}, 235023
(2010)
\end{botherref}
\endbibitem

\bibitem{color}
\begin{botherref}
\oauthor{\bsnm{Gurau}, \binits{R.}}:
Colored group field theory.
arXiv:0907.2582
(2009)
\end{botherref}
\endbibitem

\bibitem{tensor}
\begin{botherref}
\oauthor{\bsnm{Oriti}, \binits{D.}}:
The quantum geometry of tensorial group field theories.
Proceedings of The XXIX International Colloquium on Group-Theoretical Methods
  in Physics, August 20-26, 2012
(2012)
\end{botherref}
\endbibitem

\bibitem{universality}
\begin{botherref}
\oauthor{\bsnm{Gurau}, \binits{R.}}:
Universality for random tensors.
arxiv:1111.0519
(2011)
\end{botherref}
\endbibitem

\bibitem{sefu3}
\begin{botherref}
\oauthor{\bsnm{Geloun}, \binits{J.B.}},
\oauthor{\bsnm{Krajewski}, \binits{T.}},
\oauthor{\bsnm{Magnen}, \binits{J.}},
\oauthor{\bsnm{Rivasseau}, \binits{V.}}:
Linearized group field theory and power counting theorems.
Class.\ Quant.\ Grav.\ {\bf 27}, 155012; arXiv:1002.3592
(2010)
\end{botherref}
\endbibitem

\bibitem{FreiGurOriti}
\begin{botherref}
\oauthor{\bsnm{Freidel}, \binits{L.}},
\oauthor{\bsnm{Gurau}, \binits{R.}},
\oauthor{\bsnm{Oriti}, \binits{D.}}:
Group field theory renormalization - the 3d case: power counting of
  divergences.
Phys.\ Rev.\ D {\bf 80}, 044007; arXiv:0905.3772
(2009)
\end{botherref}
\endbibitem

\bibitem{sefu1}
\begin{botherref}
\oauthor{\bsnm{Magnen}, \binits{J.}},
\oauthor{\bsnm{Noui}, \binits{K.}},
\oauthor{\bsnm{Rivasseau}, \binits{V.}},
\oauthor{\bsnm{Smerlak}, \binits{M.}}:
Scaling behaviour of three-dimensional group field theory.
Class.\ Quant.\ Grav.\ {\bf 26}, 185012
(2000)
\end{botherref}
\endbibitem

\bibitem{smerlakbom}
\begin{botherref}
\oauthor{\bsnm{Bonzom}, \binits{V.}},
\oauthor{\bsnm{Smerlak}, \binits{M.}}:
Bubble divergences from cellular cohomology.
Lett.\ Math.\ Phys.\ {\bf 93}, 295
(2010)
\end{botherref}
\endbibitem

\bibitem{sylv1}
\begin{botherref}
\oauthor{\bsnm{Carrozza}, \binits{S.}},
\oauthor{\bsnm{Oriti}, \binits{D..}},
\oauthor{\bsnm{Rivasseau}, \binits{V.}}:
Renormalization of an su(2) tensorial group field theory in three dimensions.
arXiv:1303.6772
(2013)
\end{botherref}
\endbibitem

\bibitem{1on}
\begin{botherref}
\oauthor{\bsnm{Gurau}, \binits{R.}}:
The 1/n expansion of colored tensor models.
Annales Henri Poincare 12:829-847
(2010)
\end{botherref}
\endbibitem

\bibitem{Pezzana}
\begin{botherref}
\oauthor{\bsnm{Pezzana}, \binits{M.}}:
Sulla struttura topologica delle varieta compatte.
Atti Sem. Mat. Fis. Univ. Modena 23, 269-277
(1974)
\end{botherref}
\endbibitem

\bibitem{Lins}
\begin{bbook}
\bauthor{\bsnm{Lins}, \binits{S.}}:
\bbtitle{Gems, Computers and Attractors for 3-Manifolds; Series on Knots and
  Everything, Vol 5}.
\bpublisher{World Scientific},
\blocation{Singapore}
(\byear{1995})
\end{bbook}
\endbibitem

\bibitem{mio}
\begin{botherref}
\oauthor{\bsnm{Caravelli}, \binits{F.}}:
A simple proof of orientability in the colored boulatov model.
SpringerPlus, 1:6
(2012)
\end{botherref}
\endbibitem

\bibitem{gurau3}
\begin{botherref}
\oauthor{\bsnm{Gurau}, \binits{R.}}:
The complete 1/n expansion of colored tensor models in arbitrary dimension.
Annales Henri Poincare 13, 399
(2012)
\end{botherref}
\endbibitem

\bibitem{gurau4}
\begin{botherref}
\oauthor{\bsnm{Bonzom}, \binits{V.}},
\oauthor{\bsnm{Gurau}, \binits{R.}},
\oauthor{\bsnm{Riello}, \binits{A.}},
\oauthor{\bsnm{Rivasseau}, \binits{V.}}:
Critical behavior of colored tensor models in the large n limit.
Nucl. Phys. B853, 174-195
(2011)
\end{botherref}
\endbibitem

\bibitem{sylv3}
\begin{botherref}
\oauthor{\bsnm{Carrozza}, \binits{S.}},
\oauthor{\bsnm{Oriti}, \binits{D..}}:
Bubbles and jackets: new scaling bounds in topological group field theories.
HEP, Volume 2012, N 6 , 92
(2012)
\end{botherref}
\endbibitem

\bibitem{sylv4}
\begin{botherref}
\oauthor{\bsnm{Carrozza}, \binits{S.}},
\oauthor{\bsnm{Oriti}, \binits{D..}},
\oauthor{\bsnm{Rivasseau}, \binits{V.}}:
Renormalization of tensorial group field theories: Abelian u(1) models in four
  dimensions.
arXiv:1207.6734
(2012)
\end{botherref}
\endbibitem

\bibitem{jimmy}
\begin{botherref}
\oauthor{\bsnm{Ryan}, \binits{J.P.}}:
Tensor models and embedded riemann surfaces.
Phys. Rev. D 85 024010
(2012)
\end{botherref}
\endbibitem

\bibitem{geloun}
\begin{botherref}
\oauthor{\bsnm{Geloun}, \binits{J.B.}},
\oauthor{\bsnm{Rivasseau}, \binits{V.}}:
A renormalizable 4-dimensional tensor field theory.
Comm. Math. Phys. V 318 1
(2011)
\end{botherref}
\endbibitem

\bibitem{geloun2}
\begin{botherref}
\oauthor{\bsnm{Geloun}, \binits{J.B.}}:
Renormalizable models in rank $d\geq 2$ tensorial group field theory.
arXiv:1306.1201
(2013)
\end{botherref}
\endbibitem

\bibitem{rivasseau}
\begin{botherref}
\oauthor{\bsnm{Rivasseau}, \binits{V.}}:
Spheres are rare.
arXiv:1303.7371
(2013)
\end{botherref}
\endbibitem

\bibitem{gurau2}
\begin{botherref}
\oauthor{\bsnm{Gurau}, \binits{R.}}:
The 1/n expansion of tensor models beyond perturbation theory.
arXiv:1304.2666
(2013)
\end{botherref}
\endbibitem

\bibitem{knemiln}
\begin{botherref}
\oauthor{\bsnm{Milnor}, \binits{J.}}:
A unique decomposition theorem for 3-manifolds.
A. J. of Math. 84 1-7
(1962)
\end{botherref}
\endbibitem

\bibitem{GFT}
\begin{botherref}
\oauthor{\bsnm{Boulatov}, \binits{D.V.}}:
A model of three-dimensional lattice gravity.
Mod.\ Phys.\ Lett.\ A 7, 1629
(1992)
\end{botherref}
\endbibitem

\bibitem{surveyger}
\begin{botherref}
\oauthor{\bsnm{Ferri}, \binits{M.}},
\oauthor{\bsnm{Gagliardi}, \binits{C.}},
\oauthor{\bsnm{Grasselli}, \binits{L.}}:
Crystallisation moves.
A graph-theoretical representation of PL-manifolds - A survey on
  crystallizations
(1986)
\end{botherref}
\endbibitem

\bibitem{verticesbound}
\begin{botherref}
\oauthor{\bsnm{J.~B.~Geloun}, \binits{V.R.} \bsuffix{J.~Magnen}}:
Bosonic colored group field theory.
Eur.Phys.J.C70:1119-1130
(2010)
\end{botherref}
\endbibitem

\bibitem{handle1}
\begin{botherref}
\oauthor{\bsnm{Gagliardi}, \binits{C.}}:
Recognizing a 3-dimensional handle among 4-coloured graphs.
Ricerche Mat. 31 389-404
(1982)
\end{botherref}
\endbibitem

\bibitem{handle2}
\begin{botherref}
\oauthor{\bsnm{Lins}, \binits{S.}}:
A simple proof of gagliardi's handle recognition theorem.
Discrete Mathematics 57, 253-260
(1985)
\end{botherref}
\endbibitem

\end{thebibliography}







\end{backmatter}
\end{document}